\documentclass[
reprint,
superscriptaddress,
nofootinbib,
amsmath,amssymb,
aps,
prxquantum,
noeprint,
longbibliography
]{revtex4-2}

\usepackage{graphicx}
\usepackage[english]{babel}
\usepackage{adjustbox}
\usepackage{appendix}
\usepackage{amsmath}
\usepackage{braket}
\usepackage{dcolumn}
\usepackage{bm}
\usepackage{color}
\usepackage{booktabs,eqparbox}
\usepackage{dsfont}
\usepackage{placeins}
\usepackage{paralist}
\usepackage{comment}
\usepackage{multirow}
\usepackage{pgfplots}
\usepackage{import}
\usepackage{lipsum}
\usepackage{tikz}
\usepackage{csquotes}
\usetikzlibrary{shapes,arrows,fit, backgrounds, quantikz2}
\definecolor{bluempl}{RGB}{46,126,187}
\usepackage[colorlinks=true, allcolors=bluempl]{hyperref}
\usepackage{cleveref}
\usepackage{textgreek}
\usepackage{siunitx}
\usepackage[nolist,nohyperlinks]{acronym}
\usepackage{algpseudocode}
\DeclareUnicodeCharacter{2212}{−}
\usepgfplotslibrary{groupplots,dateplot}
\usetikzlibrary{patterns,shapes.arrows}
\pgfplotsset{compat=newest}

\makeatletter
\let\oldtheequation\theequation
\renewcommand\tagform@[1]{\maketag@@@{\ignorespaces#1\unskip\@@italiccorr}}
\renewcommand\theequation{(\oldtheequation)}
\makeatother

\definecolor{myRed}{HTML}{A3061E}
\definecolor{myBlue2} {RGB} {0,63,119}
\definecolor{myblue}{RGB}{70, 130, 180}
\definecolor{myYellow} {cmy} {0,0.263,0.741}
\definecolor{myGreen}{HTML}{0B6E4F}
\definecolor{FAU}{RGB}{0,56,101}
\definecolor{gray75}{gray}{0.75}
\definecolor{bluempl}{RGB}{46,126,187}

\newcommand{\imag}{\mathrm{i}}

\newcommand{\lp}{\left(} 
\newcommand{\rp}{\right)}

\newcommand{\mycomment}[1]{}

\newcommand{\GS}{\ket{\tilde{\Psi}_{\mathrm{GS}}}}
\newcommand{\GSex}{\ket{\Psi_{\mathrm{GS}}}}
\newcommand{\GSwf}{\tilde{\Psi}_{\mathrm{GS}}}
\newcommand{\exGSE}{E_{\mathrm{GS}}}
\newcommand{\subGSE}{E_{\mathrm{GS},\mathcal{B}}}

\newcommand{\Ns}{N_{\mathrm{S}}^{\mathrm{AA,it}}}
\newcommand{\Nsdir}{N_{\mathrm{S}}^{\mathrm{dir}}}

\newcommand{\Uprep}{\tilde{U}_{\mathrm{GS}}}
\newcommand{\Ft}{\mathcal{F}_{\mathrm{T}}}
\newcommand{\sel}{\mathrm{SEL}}
\newcommand{\prep}{\mathrm{PREP}}
\newcommand{\Qtot}{Q_\mathrm{tot}}

\newcounter{algorithm}
\newcommand{\Eq}{Eq.}
\newcommand{\Fig}{Fig.}

\addto\extrasenglish{%
}

\addto\extrasenglish{%
}

\graphicspath{{Figures/}}

\begin{document}
\title{Sample-Based Quantum Diagonalization with Amplitude Amplification}

\author{Nina Stockinger }
\email{nina.stockinger@fau.de}
\affiliation{Department of Physics, Friedrich-Alexander Universität Erlangen-Nürnberg, Erlangen, Germany}

\author{Ludwig Nützel }
\affiliation{Department of Physics, Friedrich-Alexander Universität Erlangen-Nürnberg, Erlangen, Germany}

\author{Michael J. Hartmann }
\email{michael.j.hartmann@fau.de}
\affiliation{Department of Physics, Friedrich-Alexander Universität Erlangen-Nürnberg, Erlangen, Germany}
\affiliation{Max Planck Institute for the Science of Light, Staudtstraße 2, 91058 Erlangen, Germany}
\affiliation{Quint Computing GmbH, Erwin-Rommel-Str. 1, 91058 Erlangen, Germany}

\begin{abstract}
   {
    Recently, sample-based quantum diagonalization (SQD) has emerged as a promising approach to compute ground and excited states of problem Hamiltonians.
This method classically diagonalizes a Hamiltonian in a subspace that is spanned by samples obtained from a quantum computer. 
However, by its nature, SQD suffers from a fundamental sampling problem, as some basis states that are required for a targeted accuracy may only be sampled extremely rarely.
To alleviate this limitation, we introduce the SQD-AA algorithm that combines SQD with amplitude amplification (AA).
\mbox{SQD-AA} uses AA to sequentially reduce probabilities of already measured bitstrings, thus making the observation of new ones more likely.
We observe a reduction in the total query complexity of more than a factor of 100 for algebraically and exponentially decaying model distributions, and analytically show a quadratic advantage for the latter.
Moreover, we evaluate real molecules in an early fault-tolerant scenario and compare SQD-AA to SQD and iterative quantum phase estimation (iQPE). 
For almost all considered examples, we observe the lowest total number of $T$-gates for SQD-AA while only requiring circuits that are 2--4 orders of magnitude shallower than those needed for iQPE.
Given this substantial reduction in circuit depth compared to iQPE while saving 2 orders of magnitude in total runtime compared to SQD, we expect a significant regime in early fault-tolerance where SQD-AA runs feasibly, but iQPE circuits are too deep to execute confidently.

   }
\end{abstract}

\maketitle

\section{Introduction}
Simulation of the electronic structure problem is widely regarded as one of the most promising applications of quantum computing, since molecular systems are inherently quantum mechanical and computing them often requires exponential resources on classical computers \cite{weidman_quantum_2024}. In particular, ground and low-lying excited state energies are of central interest, since they largely determine molecular stability, chemical reactivity, and spectroscopic properties \cite{bauer_quantum_2020}. For fault-tolerant architectures, the electronic structure problem can be solved with \ac{QPE}, likely offering polynomial and, for certain systems, eventually exponential speedups over classical approaches \cite{lee_evaluating_2023, nielsen_quantum_2010}. However, despite steady progress, there remains a substantial gap between current \ac{NISQ} devices and \ac{FASQ} machines \cite{eisert_mind_2025}.

For early \ac{FASQ}, the recently proposed quantum-centric computing is among the most promising approaches \cite{robledo-moreno_chemistry_2025}.
Within this framework, a quantum computer is embedded in \ac{HPC} to leverage the advantages of both methods. To determine \acp{GSE} of a Hamiltonian in quantum-centric computing, \ac{QSCI} \cite{kanno_quantum-selected_2023} and its variant \ac{SQD} \cite{robledo-moreno_chemistry_2025} have been introduced. Here, a Hamiltonian is diagonalized classically in a subspace determined by quantum samples. The main advantage is that a quantum computer may be used to prepare classically intractable states, whereas effects of circuit and shot noise are reduced by classical diagonalization. Shot noise is also a limiting factor when directly measuring expectation values to chemical accuracy, which requires millions of single-shot Pauli measurements at any system size \cite{gonthier_measurements_2022}. 

\ac{SQD} has also been extended to the calculation of low-lying excited states and combined with various classical methods such as selected configuration interaction, auxiliary-field quantum Monte Carlo, machine learning, or density matrix embedding theory \cite{barison_quantum-centric_2025, yoshida_auxiliary-field_2025, danilov_enhancing_2025, chen_neural_2025, cantori_adaptive-basis_2025,  patra_physics-informed_2026, patra_quantum_2026}. The different approaches have been employed to solve various molecules, metal clusters, and proteins up to 77 qubits \cite{sugisaki_hamiltonian_2025, piccinelli_quantum_2025, robledo-moreno_chemistry_2025, nutzel_solving_2025, shajan_molecular_2026}. Furthermore, applications extend to material science, for instance, to calculate band gaps, simulate battery materials, or solve molecular systems in implicit and explicit solvents \cite{duriez_computing_2025, barroca_surface_2025, kaliakin_implicit_2025, bazayeva_quantum-centric_2025}. Most commonly, a classically pre-optimized \ac{LUCJ} ansatz \cite{robledo-moreno_chemistry_2025, duriez_computing_2025, barroca_surface_2025} or time-evolution circuits \cite{piccinelli_quantum_2025, yu_quantum-centric_2025,rosanowski_sample-based_2025, sugisaki_hamiltonian_2025} are used to prepare the initial state from which bitstrings are sampled.

While promising classical sampling methods such as selected configuration interaction (SCI) and heat-bath CI (HCI) exist, recent works provide evidence that sampling from quantum circuits can yield more compact subspace expansions than their classical counterparts. For example, in the stretched-bond regime of N$_2$, quantum approaches have been shown to reduce the required subspace size by 50\,\% compared to HCI \cite{kemmoku2026generative}, indicating that static correlation hinders classical SCI methods from discovering significant determinants. Additionally, a combination of \ac{SQD} and HCI was shown to produce up to 39\,\% more compact wave functions for N$_2$ and a 76\,\% smaller subspace dimension for [4Fe-4S] than bare HCI \cite{mcfarthing2026noise}. These are strong indications that quantum SCI approaches remain viable alternatives where classical methods break down.

Yet, one of the main challenges in \ac{SQD} and other sampling-based methods is that some basis states have significantly higher probabilities than others that are also required for a target accuracy. For molecules with single-reference character, this is the \ac{HF} state; however, also for systems with multi-reference character such as Fe(III)-NTA, one or a few basis states can be dominant \cite{nutzel_solving_2025}. Moreover, even if systems do not exhibit strong multi-reference character, the exponentially growing tail of minor configurations is important to capture dynamical correlations \cite{morchen2024classification, reinholdt_critical_2025}. It follows that dominant basis states are measured very frequently, while sub-dominant basis states, which are also required for reaching the desired energy accuracy, are hardly measured at all.  This imbalance of the bitstring distribution results in a substantial measurement overhead which significantly limits the efficiency of \ac{QSCI} and \ac{SQD} \cite{reinholdt_critical_2025}.

Ideally, each basis state would be measured only once. This could be achieved if, after each single-shot measurement, the prepared quantum state would be manipulated in a way that the measured bitstring no longer contributes to the quantum state. 
Here, we introduce an algorithm that uses \ac{AA} \cite{brassard_quantum_2002} to achieve this functionality. \ac{AA} can rotate an initial state close to a desired target state via a sequence of rotations, where the amplitudes of dominant bitstrings are reduced to zero. For this procedure, we only require approximate knowledge of the probabilities of the bitstrings that are to be reduced. We therefore combine \ac{SQD} and \ac{AA} by sequentially reducing the probabilities of already measured bitstrings to obtain an algorithm that we coin \ac{SQD-AA}, which beats \ac{SQD}\footnote{We avoid the term \ac{QSCI} here because we apply \ac{SQD} beyond the scope of quantum chemistry Hamiltonians in a Slater determinant basis, where the `configuration interaction' terminology is strictly applicable.} in runtime by orders of magnitude. Since \ac{AA} preserves the relative probabilities of the bitstrings that have not yet been measured, \ac{SQD-AA} samples the same subspace as \ac{SQD}, such that the subsequent classical diagonalization is unchanged.

We analyze the algorithm's performance for algebraically and exponentially decaying model distributions. For the total query complexity as a measure for the runtime, we show that \ac{SQD-AA} achieves a quadratic advantage for the exponentially decaying, and a reduction of more than 2 orders of magnitude compared to \ac{SQD} for both distributions.

As a promising field for future applications, we further test \ac{SQD-AA} for various real quantum chemical systems.
In this context, we also provide a proof-of-principle that \ac{ASP} can serve as a scalable alternative to the \ac{UCJ} ansatz for initial state preparation.
As the depth of the circuits for \ac{AA} requires (early) fault-tolerant machines, we also compare our approach to \ac{iQPE}, which is considered among the most efficient algorithms for determining \acp{GSE} on early fault-tolerant devices \cite{martyn_grand_2021}.
A general observation is that compared to \ac{iQPE}, the deepest circuits that are executed are several orders of magnitude shallower for \ac{SQD-AA} and \ac{SQD}.
Thus, when only a limited number of logical $T$-gates can be executed, we expect an area between \ac{NISQ} and \ac{FASQ} where sample-based diagonalization methods can run, while circuits are too deep for \ac{iQPE}.
Further, comparing our \ac{SQD-AA} method to \ac{SQD}, we are able to reduce the total $T$-complexity by roughly one order of magnitude. This is caused by a reduction of the number of shots by up to a factor of 65. Therefore, our algorithm is especially useful when performing many shots is time-consuming, as is the case for trapped-ion or neutral atom quantum computers.

\section{Methods\label{sec:methods}}
Before introducing our algorithm we briefly review the essentials of \ac{SQD} and \ac{AA} to provide the necessary background for the subsequent description of \ac{SQD-AA}.

\subsection{Sample-Based Quantum Diagonalization}\label{sec:sqd}
In \ac{SQD} the eigenvalue problem is solved classically in a subspace based on quantum samples \cite{robledo-moreno_chemistry_2025, kanno_quantum-selected_2023}. Assuming that an approximate ground state $\GS$ can be prepared on a quantum computer, the state is measured $\Nsdir$ times in the computational basis, yielding a set $\mathcal{B}$ of distinct bitstrings $z_i$ with probabilities $p_i = |\langle z_i  \GS |^2$. The Hamiltonian $H$ projected onto this subspace reads
\begin{align}
    H_{\mathcal{B}} = \sum_{{z_i, z_j} \in \mathcal{B}} |z_i \rangle \langle z_i | H | z_j \rangle \langle z_j|.
\end{align}
The \ac{GSE} in the subspace, $\subGSE$, forms an upper bound to the exact GSE, $\exGSE \leq \subGSE$, according to the eigenvalue interlacing theorem \cite{kanno_quantum-selected_2023,noauthor_cauchys_nodate}. In this study, we employ the SQD method but omit the error mitigation via self-consistent configuration recovery~\cite{robledo-moreno_chemistry_2025}, as we assume an early fault-tolerant regime.

For systems where subspaces become too large to solve classically, one can divide $\mathcal{B}$ into batches, perform parallel diagonalizations of the respective subspace Hamiltonians and select the lowest subspace energy \cite{robledo-moreno_chemistry_2025}. The method gives good estimates of the \ac{GSE} if the ground (and prepared) state is sufficiently concentrated, i.e., a polynomial number of bitstrings is sufficient to determine the \ac{GSE} within a target accuracy \cite{yu_quantum-centric_2025}. As pointed out in the introduction, however, often some basis states are dominant, resulting in a high measurement overhead \cite{reinholdt_critical_2025}.

To quantify this sampling challenge, Reinholdt et al. \cite{reinholdt_critical_2025} introduced the ratio $N_\mathrm{bs}/N_\mathrm{shots}$, where $N_\mathrm{bs}$ is the number of unique bitstrings, and $N_{\mathrm{shots}}$ is the total number of shots. For N$_2$, using 10$^3$ shots, this ratio is roughly 0.1, which means that on average 0.1 new bitstrings are discovered per shot. However, when increasing the total number of shots (which one would do if more unique bitstrings are required to reach the desired accuracy), the ratio is reduced to 0.01 for 10$^6$ shots, and 0.0005 for 10$^9$ shots. Due to their large coefficients, the already measured bitstrings are sampled repeatedly, and many more samples are needed to uncover less probable bitstrings. To alleviate this sampling problem, we aim to reduce the probabilities of already measured bitstrings with \ac{AA} \cite{brassard_quantum_2002}.

\subsection{Amplitude Amplification}\label{sec:aa}
\begin{figure}
  \centering
  \includegraphics[width=5cm]{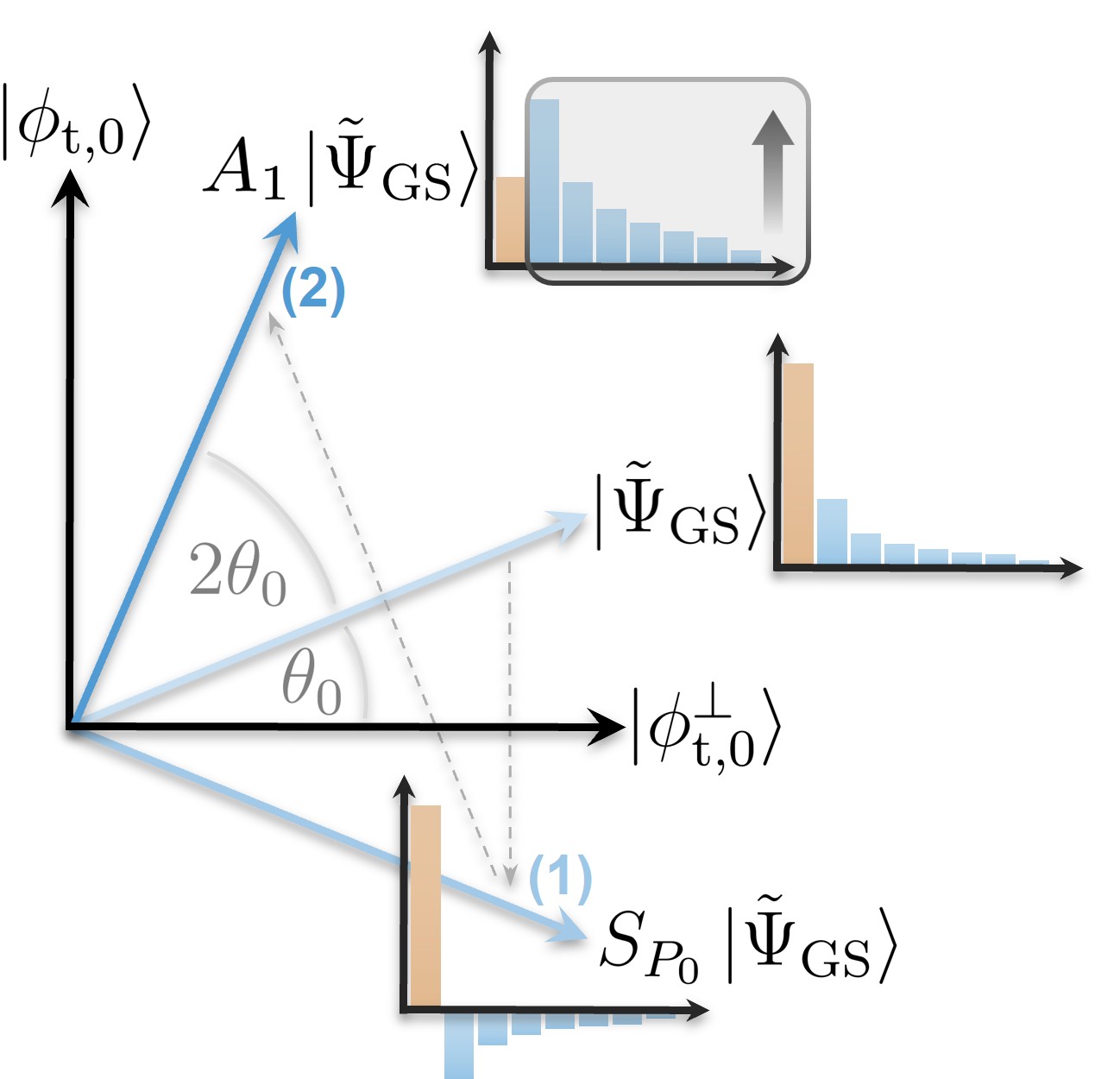}
  \caption{\textbf{Visualization of one step of AA inside a circle where the amplitude of the state marked in orange is reduced.} The bars represent real amplitudes of computational basis states. First, the phases of all basis states but the orange bitstring, $z_0$, are inverted via $S_{P_0}$ (1), followed by a reflection about $\GS$ via $S_{\GSwf}$ (2) where $A_1 = -S_{\GSwf}S_{P_0}$. $A_1$ rotates $\GS$ toward $\ket{\phi_{\mathrm{t},0}}$ by an angle of $2\theta_0$ which reduces the amplitude of the orange basis state while increasing all others on average. Based on \cite[p.253]{nielsen_quantum_2010}}
  \label{fig:aa_sub}
\end{figure}

Our goal is to rotate the prepared state $\GS = \sum_i c_i \ket{z_i}$ to a target state  $\ket{\phi_{\mathrm{t},k}}$ where the probabilities of a set of already measured bitstrings $\mathcal{S}_k = \{z_0, \ldots, z_k \}$ are reduced. For that, we express $\GS$ in a two-dimensional basis
\begin{align}
    \GS = \cos(\theta_k) \ket{\phi^\perp_{\mathrm{t},k}} + \sin(\theta_k) \ket{\phi_{\mathrm{t},k}},
\end{align}
consisting of the target state
\begin{align}
    \ket{\phi_{\mathrm{t},k}} = \frac{1}{\sqrt{1-R_k}}  \sum_{i \notin \mathcal{S}_k} c_i \ket{z_i}
    \label{eq:tarstate}
\end{align} 
and the orthogonal  $\ket{\phi^\perp_{\mathrm{t},k}} = \frac{1}{\sqrt{R_k}}  \sum_{i \in \mathcal{S}_k} c_i \ket{z_i}$ with $R_k= \sum_{i \in \mathcal{S}_k} |c_i|^2$ and where 
\begin{align}
    \theta_k = \arccos\left(\sqrt{R_k} \right).
    \label{eq:theta_via_probs}
\end{align}
To generate a rotation toward $\ket{\phi_{\mathrm{t},k}}$ we first reflect about $\ket{\phi^\perp_{\mathrm{t},k}}$ via $S_{P_k} = -(\mathbb{I} - 2P_k)$ with \mbox{$P_k = \sum_{i \in \mathcal{S}_k} |z_i\rangle \langle z_i|$}.
This is followed by a reflection about $\GS$ via \mbox{$S_{\GSwf} =  \mathbb{I} - 2 |\tilde{\Psi}_{\mathrm{GS}} \rangle \langle \tilde{\Psi}_{\mathrm{GS}}|.$}
In total, the two reflections $A_{k+1} =-S_{\GSwf}S_{P_k}$ rotate the state toward $\ket{\phi_{\mathrm{t},k}}$ by an angle of $2\theta_k$, as can be seen in \Fig \ \ref{fig:aa_sub}. As a consequence of the above operations, the amplitudes of the bitstrings in $\mathcal{S}_k$ are reduced, while all others are increased on average.

To approach the target state, the procedure is repeated $s_{k+1}$ times,
\begin{align}
     A_{k+1}^{s_{k+1}}\GS &=  \cos((2s_{k+1}+1)\theta_k)  \ket{\phi^\perp_{\mathrm{t},k}} \nonumber \\& \quad + \sin((2s_{k+1}+1)\theta_k)  \ket{\phi_{\mathrm{t},k}}.
\end{align}
Ideally, one would choose
\begin{align}
    s_{k+1} = \left\lfloor \frac{\pi}{4 \theta_k} \right \rfloor 
    \label{eq:steps}
\end{align}
such that $\sin((2s_{k+1}+1)\theta_k)$ is close to 1 and probabilities of unwanted basis states are vanishing. Note that $s_0=0$, i.e., in this case $\GS$ is prepared. To determine the ideal number of steps $s_{k+1}$, the probabilities associated with the bitstrings in $\mathcal{S}_k$ need to be determined to sufficient precision (see \Eq \ \eqref{eq:theta_via_probs}). An inaccurate estimate of $s_{k+1}$ could result in an over-rotation, meaning that the probabilities of the basis states in $\mathcal{S}_k$ are increased again. This problem could be avoided using the fixed-point version of AA, at the cost of increasing the optimal $s_{k+1}$ \cite{yoder_fixed-point_2014}. For a comparison of \ac{AA} and fixed-point \ac{AA}, we refer to Appendix \ref{subsec:aa_fpaa}. In the following, we use the standard version of \ac{AA}, as the higher optimal values of $s_{k+1}$ for fixed-point \ac{AA} would often increase the total runtime of the algorithm.

We emphasize that \ac{AA} does not alter the relative weights of the not yet sampled bitstrings, such that \ac{SQD-AA} identifies the same subspace as \ac{SQD}. Consequently, the compactness of the subspace and the cost of the classical diagonalization are identical for \ac{SQD} and \ac{SQD-AA}, and all statements on the compactness of \ac{SQD} subspaces directly carry over to \ac{SQD-AA}.

\begin{figure*}
  \centering
  \includegraphics[width=\textwidth]{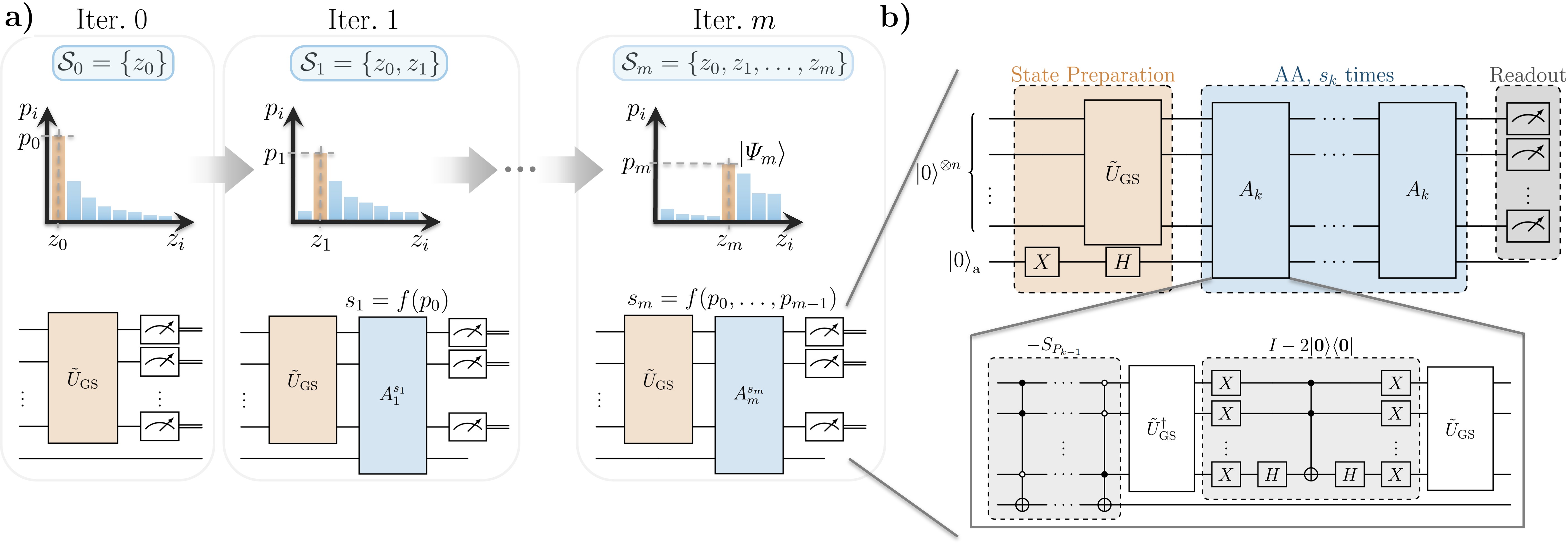}
  \caption{\textbf{Sketch of SQD-AA. a)} First, an approximate ground state $\GS = \ket{\Psi_0}$ is prepared and measured in the computational basis, yielding the most probable bitstring $z_0$ and its approximate probability $p_0$. The steps to reduce the probability of $z_0$ are determined as function of $p_0$, $s_1=f(p_0)$, via Eqs.~\eqref{eq:theta_via_probs} and \eqref{eq:steps}. Applying $A_1$ $s_1$ times to $\GS$ results in the next state where $z_1$ has the highest probability. The procedure is repeated until either the energy converges or the distribution becomes too flat. In the second case, the final state $\ket{\Psi_{m}}$ is measured until $\Delta E_{\mathcal{B}}$ (\Eq~\eqref{eq:conv_energy_sqdaa}) is below a threshold. \textbf{b)} Schematic of the circuits. First, $\Uprep$ is applied to $n$ qubits and an ancilla is prepared in the $\ket{-}$ state. Then, $A_k$ is executed $s_k$ times and the system qubits are measured in the computational basis. Each $A_k$ consists of a reflection $-S_{P_{k-1}}$, inverting the phases of all basis states in $\mathcal{S}_{k-1}$, followed by a reflection about the ground state $S_{\GSwf}$ which is implemented via application of $\Uprep$ and its Hermitian adjoint to $\mathbb{I} - 2|\mathbf{0}\rangle \langle \mathbf{0}|$.}
  \label{fig:alg}
\end{figure*}

The circuit that implements \ac{AA} is shown in \Fig~\ref{fig:alg}~b) \cite[p.248--256]{nielsen_quantum_2010}. First, $\Uprep$ is applied and an ancilla is prepared in the $\ket{-}$ state. Subsequently, $A_{k+1}$ is applied $s_{k+1}$ times. Within $A_{k+1}$, $-S_{P_k}$ flips the phases of all bitstrings in $\mathcal{S}_k$ which is achieved via multi-qubit CNOT gates controlled by the respective basis states acting on the ancilla. These multi-controlled CNOT gates can be decomposed into Clifford and $T$-gates with linear complexity in the number of qubits \cite{nakanishi_systematic_2024} and therefore do not contribute significantly to the overall cost. That is, we require $N_{T,\mathrm{C}^n\mathrm{NOT}} = 4n - 6$ $T$-gates per C$^n$NOT gate, and have $((k+1) + 1)s_{k+1}$ multi-qubit CNOT gates per iteration $k$. The reflection $S_{\GSwf}$ is implemented via the ground-state preparation unitary $\Uprep$ and its Hermitian adjoint acting on $\mathbb{I} - 2\ket{\mathbf{0}}\langle \mathbf{0}|$. Assuming that $N_{\mathrm{T}, \Uprep} \gg 4n-6$, the dominant cost of the circuit arises from applying $\Uprep$ $2s_{k+1}+1$ times. Having described how to adapt probabilities with AA, we are now in a position to introduce our novel algorithm that combines \ac{SQD} with AA.

\subsection{Sample-Based Quantum Diagonalization with Amplitude Amplification}\label{sec:sqdaa}
Within \ac{SQD-AA}, we iteratively apply AA to reduce the probabilities of dominant bitstrings sequentially. First, an approximate ground state is prepared as in \ac{SQD}, $\Uprep \ket{\mathbf{0}} = \GS$. Starting with $\GS = \ket{\Psi_0}$ for $k=0$ and $s_0=0$, in each iteration $k$, the current state $\ket{\Psi_k}$ is measured $\Ns$ times in the computational basis. This allows us to obtain a dominant bitstring $z_k$ that is not yet in $\mathcal{S}_{k-1}$ and its approximate probability $p_k^{(s_k)}$ that we want to reduce in the following. For that, we always start with the initial state $\ket{\Psi_0}$ and reduce the probabilities of all selected bitstrings, $\mathcal{S}_k = \{z_j\}_{j=0}^{k}$, simultaneously. All measured bitstrings are collected in the set $\mathcal{B}$, which spans the subspace for the diagonalization (see Section~\ref{sec:sqd}). (Note that in the first iteration, the initial state is measured until at least two different bitstrings are obtained.) To determine the number of steps $s_{k+1}$ via \Eq \ \eqref{eq:theta_via_probs} and \eqref{eq:steps}, however, we need the bitstring probabilities $p_k^{(0)}$ that appear in $\ket{\Psi_0}$, and not $p_k^{(s_k)}$ from $\ket{\Psi_k}$. Therefore, we estimate the probabilities in the initial state $\ket{\Psi_0}$ recursively as
\begin{align}
    p_{k}^{(0)} = \frac{1 - \sum_{i=0}^{k-1}{p_{i}^{(0)}}}{1 - \sum_{i=0}^{k-1} p_{i}^{(s_k)}} \cdot p_{k}^{(s_k)}.
    \label{eq:prob_est}
\end{align}
Additionally, we introduce a target fidelity $\mathcal{F}_{\mathrm{T}} = |\langle \phi_{\mathrm{t},k}|A_{k+1}^{s_{k+1}} \GS|^2$ of the rotated state and the current target state $\ket{\phi_{\mathrm{t},k}}$, and determine the steps such that $\sin^2((2s_{k+1}+1)\theta_k) \approx \mathcal{F}_{\mathrm{T}}$. Note that we do not need classical representations of $\ket{\phi_{\mathrm{t},k}}$ and $\GS$ for that, but only the angle $\theta_k$ determined via \Eq \ \eqref{eq:theta_via_probs}. Setting $\mathcal{F}_{\mathrm{T}} < 1$ can reduce the probability of over-rotations. Once the number of steps is determined, the \ac{AA} unitary is applied $s_{k+1}$ times, producing the state $\ket{\Psi_{k+1}} = A_{k+1}^{s_{k+1}} \ket{\Psi_0}$ for the next iteration.

As the probabilities $p_k^{(s_k)}$ are estimated with a finite number of shots, they are subject to statistical uncertainty. This error is transferred to the estimated number of steps via \Eq \ \eqref{eq:theta_via_probs} and \eqref{eq:steps}. If the error in $s_{k+1}$ is too large, $\ket{\Psi_{k+1}}$ might not be sufficiently close to the target state $\ket{\phi_{\mathrm{t},k}}$ to measure a bitstring outside $\mathcal{S}_k$. For instance, if the number of estimated steps $s_{k+1}$ is too large, $\ket{\Psi_{k+1}}$ might be close to $\ket{\phi_{t,k}^{\perp}}$.
In this case, the number of steps, $s_{k+1}$, can be adapted heuristically. To do so, we need to determine whether the current number of steps is above or below the ideal number of steps. For that, we can reduce or increase the number of steps according to its order of magnitude and measure the new state $\ket{\Psi_{(k+1)'}} = A_{k+1}^{s_{k+1}'} |\Psi_{0} \rangle$. If we reduce the number of steps, i.e., $s_{k+1}' < s_{k+1}$, and the sum of the remaining probabilities of the bitstrings that we reduce becomes smaller, $\sum_{i=0}^{k} p_i^{(s_{k+1}')} < \sum_{i=0}^{k} p_i^{(s_{k+1})}$, the number of steps can be considered too large and we can reduce it further until we measure a bitstring outside $\mathcal{S}_k$, and vice versa. The accumulated change is then added permanently to all later step counts,  since the later rotations start from the same $\ket{\Psi_0}$ and suffer from the same misestimation.

Finally, we need to introduce a convergence criterion. As will be discussed in Section \ref{subsec:dif_model_dists} and Appendix \ref{subsec:sample_comp_sqdaa}, amplitude reduction is only more efficient than direct measurements if the distribution is sufficiently uneven. To estimate the rate of decay of the probabilities of computational basis states, we introduce the relative difference  
\begin{align}
    \Delta_{k-1,k} = 2 \, \frac{|p_{k-1}^{(0)} - p_k^{(0)}|}{p_{k-1}^{(0)} + p_{k}^{(0)}}.
    \label{eq:conv_delta}
\end{align}
Hence, if $\Delta_{k-1,k}$ is below or equal to a threshold $\tau$ and the steps to reduce the next bitstring are not equal to the previous steps, $s_{k+1} \neq s_k$, (if this would be the case, the probability of the next bitstring could be reduced with almost no additional cost, as the main cost arises from applying $\Uprep$ $(2s_k+1)$ times), we do not reduce probabilities further, but measure the current state $\ket{\Psi_k}$ until the GSE in the subspace converges, i.e., whenever newly sampled bitstrings enlarge the set from $\mathcal{B}^{\mathrm{prev}}$ to $\mathcal{B}$, 
\begin{align}
    \Delta E_{\mathcal{B}} = |E_{\mathrm{GS},\mathcal{B}^{\mathrm{prev}}} - E_{\mathrm{GS},\mathcal{B}}| \leq \epsilon.
    \label{eq:conv_energy_sqdaa}
\end{align}
The overall convergence criterion for the \ac{GSE} is also tested within each iteration, since every measurement may add new bitstrings to $\mathcal{B}$, and might be met before $\Delta_{k-1,k} \leq \tau$. Note that for the sake of comparison in the simulations within this work, where exact \acp{GSE} are known, we run \ac{SQD} and \ac{SQD-AA} until a desired energy error, $|E_{\mathrm{GS},\mathcal{B}} - \exGSE| \leq \epsilon$, is reached instead of the convergence criterion in \Eq \ \eqref{eq:conv_energy_sqdaa}.

\begin{figure}[t]
	\rule{\linewidth}{1.1pt}
	\vspace{0.1em}
	\refstepcounter{algorithm}\label{algo:sqdaa}\textbf{Algorithm~\thealgorithm:} SQD-AA
	\vspace{0.1em}
	\hrule
	\begin{algorithmic}[1]
		\State \textbf{Input:} $\Uprep$ with $\Uprep\ket{\mathbf{0}} = \ket{\Psi_0} = \GS$; $\Ns$, $\Ft$, $\tau$, $\epsilon$
		\State \textbf{Initialize:} $k \gets 0$, $s_0 \gets 0$, $\mathcal{S}_{-1} \gets \emptyset$, $\mathcal{B} \gets \emptyset$
		\Loop
		\State Measure $\ket{\Psi_k}$ $\Ns$ times \par \hspace{\algorithmicindent} $\rightarrow$ distinct bitstrings $\mathcal{M}_k$ with relative \par \hspace{\algorithmicindent} frequencies $p_i^{(s_k)}$
		\State $\mathcal{B} \gets \mathcal{B} \cup \mathcal{M}_k$, \par \hspace{\algorithmicindent} determine $E_{\mathrm{GS},\mathcal{B}}$ and $\Delta E_{\mathcal{B}}$
		\If{$\Delta E_{\mathcal{B}} \leq \epsilon$} \textbf{return} $E_{\mathrm{GS},\mathcal{B}}$
		\EndIf
		\If{$\mathcal{M}_k \setminus \mathcal{S}_{k-1} \neq \emptyset$}
		\State $z_k \gets$ most frequent bitstring in $\mathcal{M}_k \setminus \mathcal{S}_{k-1}$
		\State $\mathcal{S}_k \gets \mathcal{S}_{k-1} \cup \{z_k\}$
		\State Estimate $p_k^{(0)}$ via \Eq~\eqref{eq:prob_est} and steps \par
		\hspace{\algorithmicindent} $s_{k+1} = \left \lfloor \frac{\arcsin \left (\sqrt{\Ft}\right ) }{2 \arccos \left(\sqrt{\sum_{i=0}^k p_{i}^{(0)}} \right)} \right \rfloor$
		\If{$k \geq 1$ \textbf{and} $\Delta_{k-1,k} \leq \tau$ \textbf{and} $s_{k+1} \neq s_k$}
		\State \textbf{break} \Comment{measure $\ket{\Psi_k}$ directly}
		\EndIf
		\State $\ket{\Psi_{k+1}} = A_{k+1}^{s_{k+1}} \ket{\Psi_0}$, \quad $k \gets k + 1$
		\Else
		\State Adapt $s_k \to s_k'$, \quad $\ket{\Psi_{k}} = A_k^{s_k'} \ket{\Psi_0}$
		\EndIf
		\EndLoop
		\Repeat
		\State Measure $\ket{\Psi_k}$, add new bitstrings to $\mathcal{B}$, \par \hspace{\algorithmicindent} update $E_{\mathrm{GS},\mathcal{B}}$ and $\Delta E_{\mathcal{B}}$
		\Until{$\Delta E_{\mathcal{B}} \leq \epsilon$}
		\State \textbf{Output:} $E_{\mathrm{GS},\mathcal{B}}$ (and $\ket{\Psi_{\mathrm{GS},\mathcal{B}}}$)
	\end{algorithmic}
	\hrule
\end{figure}

The algorithm is summarized in Algorithm \ref{algo:sqdaa} and sketched in \Fig \ \ref{fig:alg} a). Here, we can see that reducing the probabilities of already measured bitstrings can significantly reduce the sample complexity, however, at the cost of deeper circuits.

\section{Results\label{sec:results}}
To gain a deeper understanding of how SQD-AA improves upon SQD, we compare both methods for different model distributions, followed by a validation on several molecules as examples.

\subsection{SQD-AA for Model Distributions}\label{subsec:dif_model_dists}
As can be seen, for example, in Ref. \cite{robledo-moreno_chemistry_2025}, \Fig \ S9 or in the Appendix, \Fig \ \ref{fig:gain_comp_hqs}, there exist electronic structure problems where the probabilities of bitstrings in the ground state decay algebraically or (piecewise) exponentially. Therefore, we consider an algebraically ($p_l \propto (l+1)^{-\gamma}$) and an exponentially ($p_l \propto e^{-\alpha l}$) decaying model distribution as example for an analytic comparison of \ac{SQD} and \ac{SQD-AA}. To compare the algorithms, we analyze the runtime required to obtain the $m$ most probable bitstrings. 

For \ac{SQD-AA}, first the probabilities of the $m^*-1$ most probable basis states are reduced sequentially in $m^*$ iterations, where $m^* \leq m$. That is, we aim to rotate the initial states 
\begin{align}
    \ket{\Psi_{\mathrm{alg}}} = \frac{1}{\sqrt{\mathcal{N}_{\mathrm{alg}}}} \sum_{l=0}^{N-1} (l+1)^{-\gamma/2} \ket{l}
\end{align}
with $\mathcal{N}_{\mathrm{alg}} = \sum_{l=0}^{N-1} (l+1)^{-\gamma}$ and
\begin{align}
    \ket{\Psi_{\mathrm{exp}}} = \frac{1}{\sqrt{\mathcal{N}_{\mathrm{exp}}}} \sum_{l=0}^{N-1} e^{-\alpha l/2} \ket{l}
\end{align}
with $\mathcal{N}_{\mathrm{exp}} = \sum_{l=0}^{N-1} e^{-\alpha l}$ to target states $\ket{\phi_{\mathrm{t},k}}$ (see \Eq \ \eqref{eq:tarstate}) for $k=0,1, \ldots,m^*-2$. Here, $N=2^n$ with the number of qubits $n$, while $\alpha$ and $\gamma$ are parameters that tune the rate of decay of the amplitudes. The number of steps to reduce the probabilities of bitstrings $\{z_i\}_{i=0}^k$ in the $k$th iteration can be estimated as
\begin{align}
    s_{k+1} &= \left \lfloor \frac{\pi}{4 \theta_k} \right \rfloor \approx \left \lfloor \frac{\pi}{4 \sin(\theta_k)} \right \rfloor  = \left \lfloor \frac{\pi}{4 \langle \tilde{\Psi}_\mathrm{GS} | \phi_{\text{t} ,k} \rangle} \right \rfloor,
\end{align}
where we use $\theta_k \ll 1$. More specifically, when calculating the overlap $\langle \tilde{\Psi}_\mathrm{GS} | \phi_{\text{t} ,k} \rangle$ we obtain
\begin{align}
    s_{k+1, \mathrm{exp}} \approx \left \lfloor \frac{\pi \sqrt{e^{\alpha (k+1)}}}{4} \right \rfloor
    \label{eq:steps_exp_dec}
\end{align}
for the exponentially decaying state and
\begin{align}
    s_{k+1, \mathrm{alg}} \approx \left \lfloor \frac{\pi \sqrt{\zeta (\gamma)}}{4 \sqrt{\zeta (\gamma) - H_{k+1}(\gamma)}} \right \rfloor
    \label{eq:steps_alg_dec}
\end{align}
for the algebraically decaying state. Here, we introduce the Riemann zeta function $\zeta(\gamma)$ as an approximation of the sum $\sum_{l=0}^{N-1} (l+1)^{-\gamma}$ and define the $k$th harmonic number of order $\gamma$, $H_k(\gamma) = \sum_{l=0}^{k-1} (l+1)^{-\gamma}$. We refer to Appendix \ref{subsec:sample_comp_sqdaa} for a detailed derivation of the results in this section.

Within each iteration of \ac{SQD-AA}, the dominant cost arises from applying the state preparation unitary $Q_k = 2 s_{k} +1$ times (see \Fig \ \ref{fig:alg} b), where we introduce the query complexity $Q_k$. Additionally, each (rotated) state $\ket{\Psi_k}$ is measured in the computational basis $\Ns$ times to determine the next number of steps $s_{k+1}$ with sufficient precision. Therefore, we introduce the total query complexity of $m^*$ iterations of AA
\begin{align}
    Q_\mathrm{tot,AA}^\mathrm{SQD-AA} = \Ns \sum_{k=0}^{m^*-1} Q_k,
\end{align}
as a measure for total runtime. As described in Section~\ref{sec:sqdaa}, amplitude reduction is only more efficient than direct sampling if the distribution is sufficiently decaying. In case the distribution becomes too flat, the state of the $(m^*-1)$th iteration, $\ket{\Psi_{m^*-1}}$, is measured directly until all $m$ bitstrings are obtained. This yields an additional contribution
\begin{align}
    Q_\mathrm{tot,dir}^\mathrm{SQD-AA} &\approx N_\mathrm{S}^{\mathrm{AA,dir}} Q_{m^*-1} \nonumber \\ &\approx \frac{1}{p_{m-1}(\Psi_{m^*-1})} \ln{\frac{m-m^*}{p_{\mathrm{fail}}}} Q_{m^*-1}
    \label{eq:qtot_sqdaa_dir}
\end{align}
that has to be added to $Q_\mathrm{tot,AA}^\mathrm{SQD-AA}$. Here, the number of shots $N_\mathrm{S}^{\mathrm{AA,dir}}$ is estimated such that the probability $p_{\mathrm{fail}}$ of not seeing one of the remaining $m-m^*$ bitstrings is upper bounded by
\begin{align}
    p_{\mathrm{fail}} &:= \sum_{k=m^*}^{m-1} (1-p_k(\Psi_{m^*-1}))^{N_\mathrm{S}^{\mathrm{AA,dir}}} \nonumber \\ &\leq \sum_{k=m^*}^{m-1} (1-p_{m-1}(\Psi_{m^*-1}))^{N_\mathrm{S}^{\mathrm{AA,dir}}}  \nonumber \\ &\leq (m-m^*) \cdot e^{-N_\mathrm{S}^{\mathrm{AA,dir}}\, p_{m-1}(\Psi_{m^*-1})}
    \label{eq:pfail}
\end{align}
where we use that the bitstrings are ordered by decreasing probability, $p_{k+1} \leq p_k$. Therefore, the total query complexity for \ac{SQD-AA} is given by
\begin{align}
    Q_\mathrm{tot}^\mathrm{SQD-AA} = Q_\mathrm{tot,AA}^\mathrm{SQD-AA} + Q_\mathrm{tot,dir}^\mathrm{SQD-AA}.
\end{align}
Note that for the algebraically decaying distribution usually $m^* \ll m$, whereas for the exponentially decaying distribution $m^* \approx m$. That is, the algebraic distribution becomes relatively flat for large values of $m$, whereas the exponentially decaying distribution is always sufficiently decaying such that AA is more efficient than direct sampling.

For bare \ac{SQD}, the total query complexity $Q_\mathrm{tot}^\mathrm{SQD}$ is equal to the total number of shots, since $\Uprep$ is applied once for each shot. As for $Q_\mathrm{tot,dir}^\mathrm{SQD-AA}$, we estimate the number of shots $\Nsdir$ to sample all important bitstrings with high probability $1-p_{\mathrm{fail}}$,
\begin{align}
    Q_\mathrm{tot}^\mathrm{SQD} &= \Nsdir \geq \frac{1}{p_{m-1}(\tilde{\Psi}_\mathrm{GS})} \ln{\frac{m}{p_{\mathrm{fail}}}}.
    \label{eq:s_exp_sqd}
\end{align}

\begin{figure}
    \centering
    \includegraphics[width=\columnwidth]{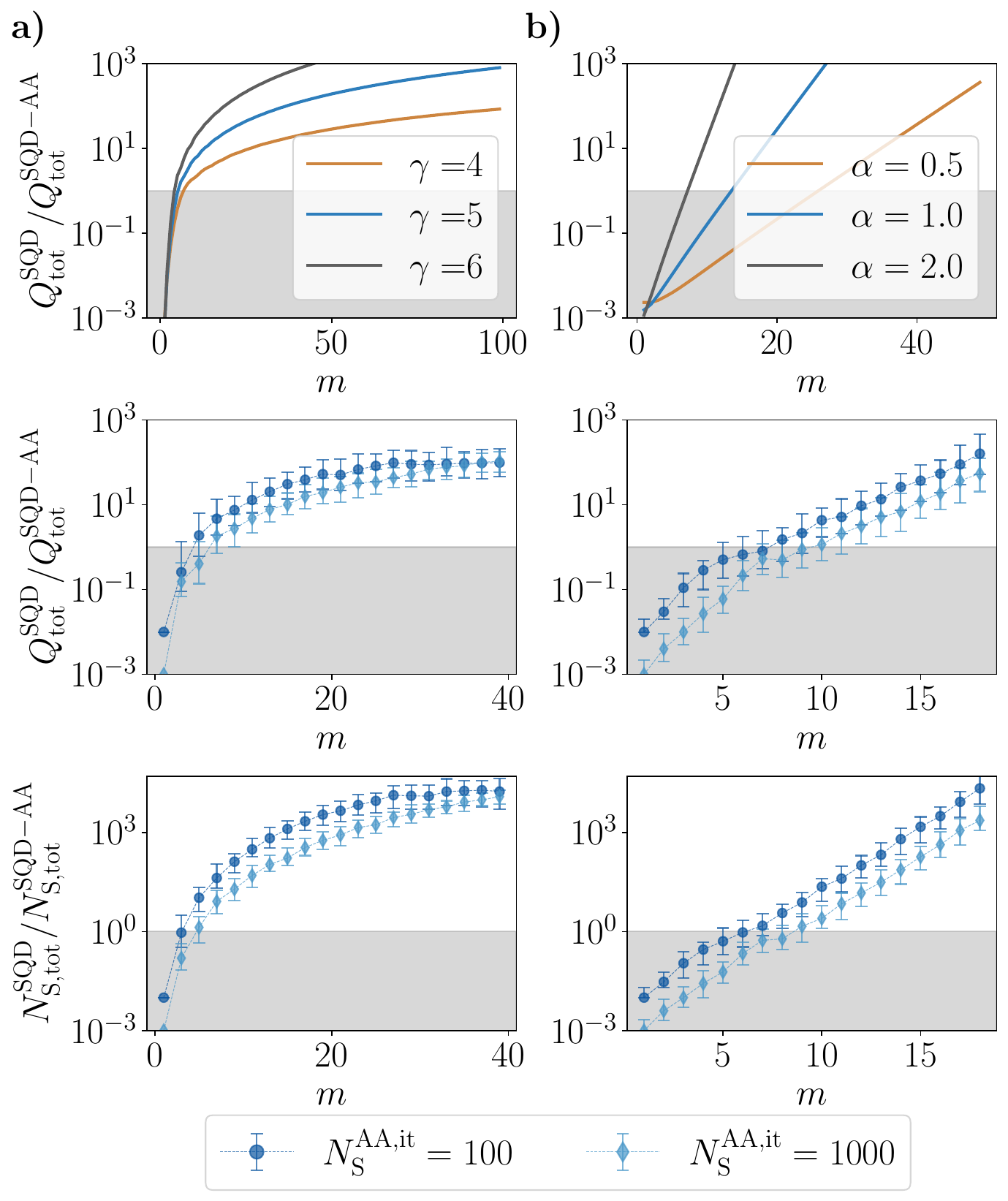}
    \caption{\textbf{Comparison of total query complexities for SQD and SQD-AA assuming an algebraically (a) and an exponentially (b) decaying distribution.} 
    In the first row, estimated ratios $Q_\mathrm{tot}^\mathrm{SQD} / Q_\mathrm{tot}^\mathrm{SQD-AA}$ are plotted for different parameters $\gamma$ or $\alpha$. Here, $\Ns=1000$, whereas the shot counts for direct measurements, $N_\mathrm{S}^{\mathrm{AA,dir}}$ and $\Nsdir$, are approximated via \Eq \ \eqref{eq:qtot_sqdaa_dir} and \eqref{eq:s_exp_sqd}, with $p_{\mathrm{fail}}=0.1$. The exact expressions for the query complexities can be found in Appendix \ref{subsec:sample_comp_sqdaa}. 
    In lower panels, we show the median reduction in $Q_\mathrm{tot}$ (second row) and $N_{\mathrm{S,tot}}$ (third row) running both algorithms with simulated measurements until all $m$ most probable bitstrings are obtained. 
    (Note that $N_\mathrm{S,tot}^\mathrm{SQD}  = \Nsdir$ and $N_\mathrm{S,tot}^\mathrm{SQD-AA} = m^* \times \Ns + N_{\mathrm{S}}^\mathrm{AA,dir}$.)
    We assume 10 qubits for the simulations and choose $\gamma=5$, $\Ft=0.8$, and $\tau=0.4$ for $\ket{\Psi_{\mathrm{alg}}}$ and $\alpha=1$, $\Ft=0.7$, and $\tau=0.3$ for $\ket{\Psi_{\mathrm{exp}}}$. Error bars indicate the 68\,\% range of 100 restarts. The gray shaded region highlights the range where no reduction in $Q_\mathrm{tot}$ or $N_{\mathrm{S,tot}}$ is achieved when using \ac{SQD-AA}. 
    }
    \label{fig:expalgcomp}
\end{figure}

Inserting corresponding quantities for the exponentially decaying distribution, which we derive in detail in Appendix~\ref{subsec:sample_comp_sqdaa}, we obtain query complexities that scale as $Q_\mathrm{tot}^\mathrm{SQD-AA}\propto \sqrt{e^{\alpha m}}$ and $Q_\mathrm{tot}^\mathrm{SQD}\propto e^{\alpha m}$. We therefore obtain a quadratic advantage in the total query complexity for SQD-AA. For the algebraically decaying distribution, the analytic expression is more complex and we do not provide it here. Instead, we plot the ratios $Q_\mathrm{tot}^\mathrm{SQD} / Q_\mathrm{tot}^\mathrm{SQD-AA}$ for both distributions in \Fig \ \ref{fig:expalgcomp} to analyze the relation between the total query complexities of \ac{SQD-AA} and \ac{SQD} in more detail.

Results for the algebraically decaying distribution are shown in \Fig \ \ref{fig:expalgcomp} a), while the ratios for the exponentially decaying distribution are plotted in \Fig \ \ref{fig:expalgcomp} b). In the upper panels of \Fig \ \ref{fig:expalgcomp}, we estimate the reduction in the total query complexity ($Q_\mathrm{tot}^\mathrm{SQD} / Q_\mathrm{tot}^\mathrm{SQD-AA}$) for different parameters $\alpha$ and $\gamma$ at 100 qubits.
Since the qubit number occurs only in the normalization factors that approach constant values with increasing $N$, the reduction in query complexity will be very similar at any system size.
For all distributions, we observe a reduction in $\Qtot$ of up to two orders of magnitude at different subspace dimensions $m$. 
In our examples, the least probable bitstrings occur with probabilities of $\sim 10^{-8}$.
Even more significant reductions are possible if one targets higher accuracies, i.e., measuring bitstrings with lower amplitudes.
For more rapidly decaying distributions, (i.e., for larger values of $\alpha$ or $\gamma$), the factor of reduction is growing faster. Of course, in that case the minimum subspace dimension required to reach a certain accuracy threshold also decreases. Nonetheless, we find that SQD-AA yields larger reduction factors for more rapidly decaying distributions.

To see how shot noise would influence the results, we run \ac{SQD-AA} for both distributions with simulated measurements and different $\Ns$ until we obtain the $m$ most probable bitstrings. 
For these simulations, we show the reduction in $\Qtot$ in the second row and the corresponding reduction in the total number of shots $N_\mathrm{S,tot}$ in the third row of \Fig \ \ref{fig:expalgcomp}.
For both states, we observe an increased reduction in $\Qtot$ when using a lower number of shots, $\Ns=100$. With this number of shots, we observe an advantage in the total query complexity using \ac{SQD-AA} for $m>3$ for the algebraically decaying and for $m> 7$ for the exponentially decaying distribution.
Moreover, we observe a reduction in $\Qtot$ of more than a factor of 100 for $m>26$ for the algebraically decaying and for $m>16$ for the exponentially decaying distribution.
This runtime reduction is caused by a reduction in the sample complexity, as can be seen in the lower panels of \Fig \ \ref{fig:expalgcomp}. Here, we observe a reduction in $N_\mathrm{S,tot}$ of up to 4 orders of magnitude for both systems. 
Therefore, \ac{SQD-AA} is particularly useful for neutral atom or trapped-ion devices, where performing many shots is expensive.  
When considering $\Qtot$, this factor of reduction is lower due to the deeper circuits; however, we still obtain a net reduction in the total runtime of more than 2 orders of magnitude.
These results offer a first insight into the advantage that can be achieved with \ac{SQD-AA}. We now explore this further for real molecules.

\subsection{Benchmarking SQD-AA for different Molecules}\label{subsec:benchmark_sqdaa}
To further investigate our approach and corroborate its usefulness, we test \ac{SQD-AA} for various molecules of interest. First, we consider cyclopentadiene, for which the spectral gap is of interest for electron spectroscopy.  Cyclopentadiene can be described by a Hamiltonian derived in \cite{shirazi_efficient_2024, hqs2026hqstage} via \ac{RPA}. This Hamiltonian consists of an active space comprising two \ac{MOs} coupled to a bath formed by the other \ac{MOs}. The number of qubits determines the truncation level, i.e., the number of environment orbitals, see Appendix \ref{subsec:molecules} for further information.

In a qubit basis, the Hamiltonian contains only a small number of Pauli terms. Therefore, we choose \ac{ASP} as a scalable method to prepare an approximate ground state $\GS$. Details of \ac{ASP} are provided in Appendix \ref{subsec:state_prep}. We then compare the resources to estimate the \acs{GSE}s of the respective Hamiltonians (i.e., for different numbers of qubits) with \ac{SQD-AA} and \ac{SQD} to chemical accuracy in the active space (i.e., an energy error of $\epsilon = 1.6\times 10^{-3}$\,Ha). 

Note that we do not include a comparison to classical selected-CI methods such as HCI. Whether sample-based methods can outperform such methods in terms of compactness or overall runtime is an open question that is independent of the sampling procedure and has been investigated elsewhere~\cite{reinholdt_critical_2025, kemmoku2026generative, mcfarthing2026noise}.

We choose the total number of $T$-gates as the quantity to measure  the effort of both methods. This is due to the fact that \ac{SQD-AA} can only run on an (early) fault-tolerant quantum computer because of the relatively deep circuits. In this case, $T$-gates are the dominant cost, as they rely on expensive magic state distillation. Therefore, $T$-complexity is often used to compare runtimes of fault-tolerant quantum algorithms \cite{kivlichan_improved_2020, babbush_encoding_2018}. (Note that for \ac{SQD-AA}, we also include the $T$-counts required to decompose the MCNOT gates of the oracles and phase gates.)

Since we require (early) fault-tolerant quantum computing, we additionally compare our method to \ac{iQPE} (see Appendix \ref{sec:appC}), which is regarded as one of the most efficient algorithms to determine \ac{GSE}s on (early fault-tolerant) quantum computers \cite{dobsicek_arbitrary_2007, lee_evaluating_2023}. Here, we choose \ac{iQPE} instead of \ac{QPE} as individual circuits are shallower and we assume early fault-tolerance where only a limited number of logical $T$-gates can be executed within sufficiently low error rates. Moreover, we give a brief comparison to other phase estimation methods in Appendix~\ref{sec:appC}. Since eigenvalues of a Hamiltonian are estimated on a quantum computer within \ac{iQPE}, the Hamiltonian must be encoded in a unitary. The most common approaches are Trotterization (see Appendix \ref{subsec:iqpe_trotter}) \cite{kivlichan_improved_2020} and Qubitization (see Appendix \ref{subsec:iqpe_qubitization}) \cite{babbush_encoding_2018}. Qubitization can yield favorable $T$-counts, especially with increasing system size, at the cost of more ancillas. In addition to briefly reviewing these methods, we describe how we obtain the respective $T$-counts for an energy error of $\epsilon = 1.6\times 10^{-3}$\,Ha in~Appendix~\ref{sec:appC}.

\begin{figure}
    \centering
    \includegraphics[width=\columnwidth]{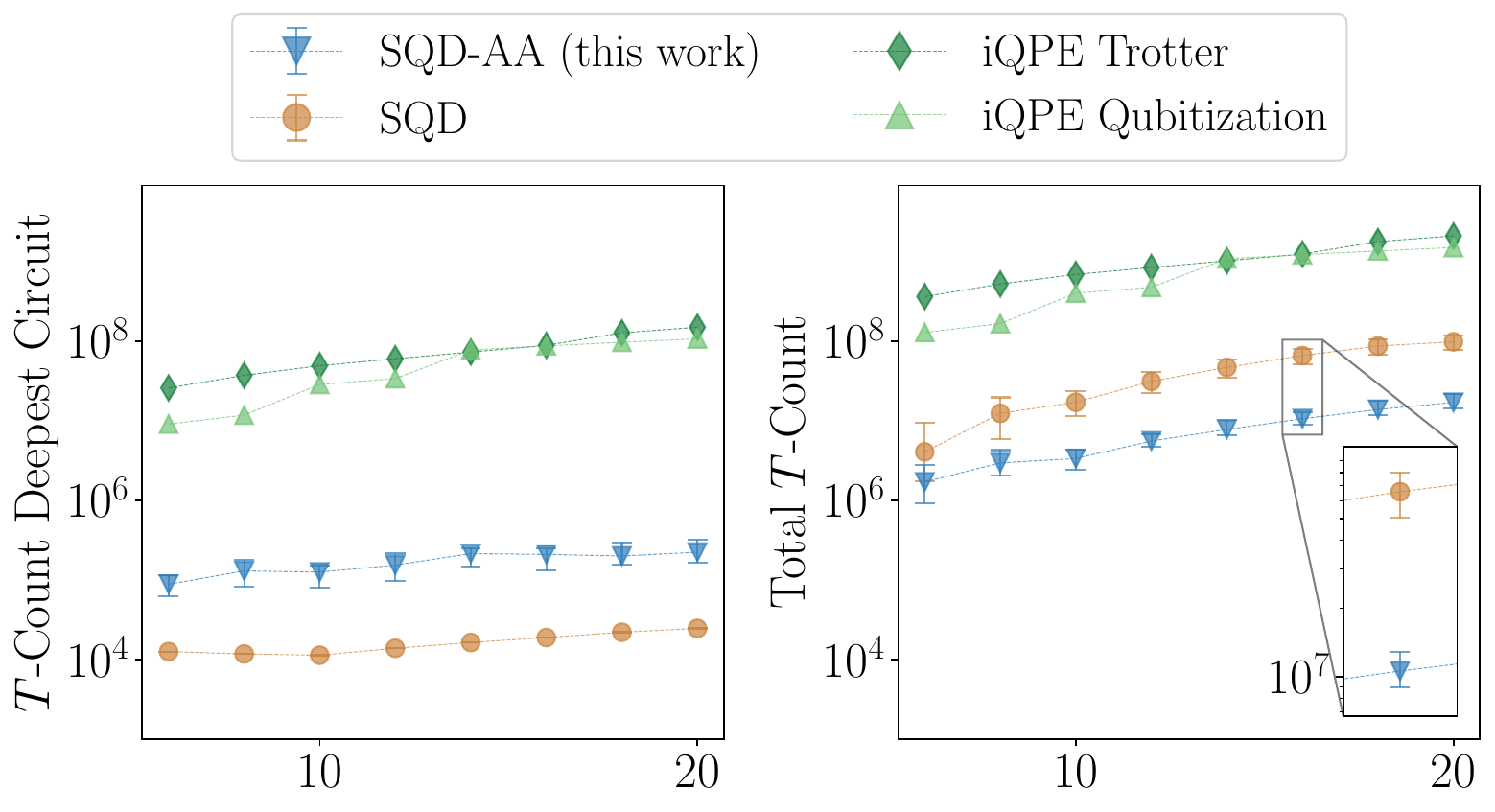}
    \includegraphics[width=\columnwidth]{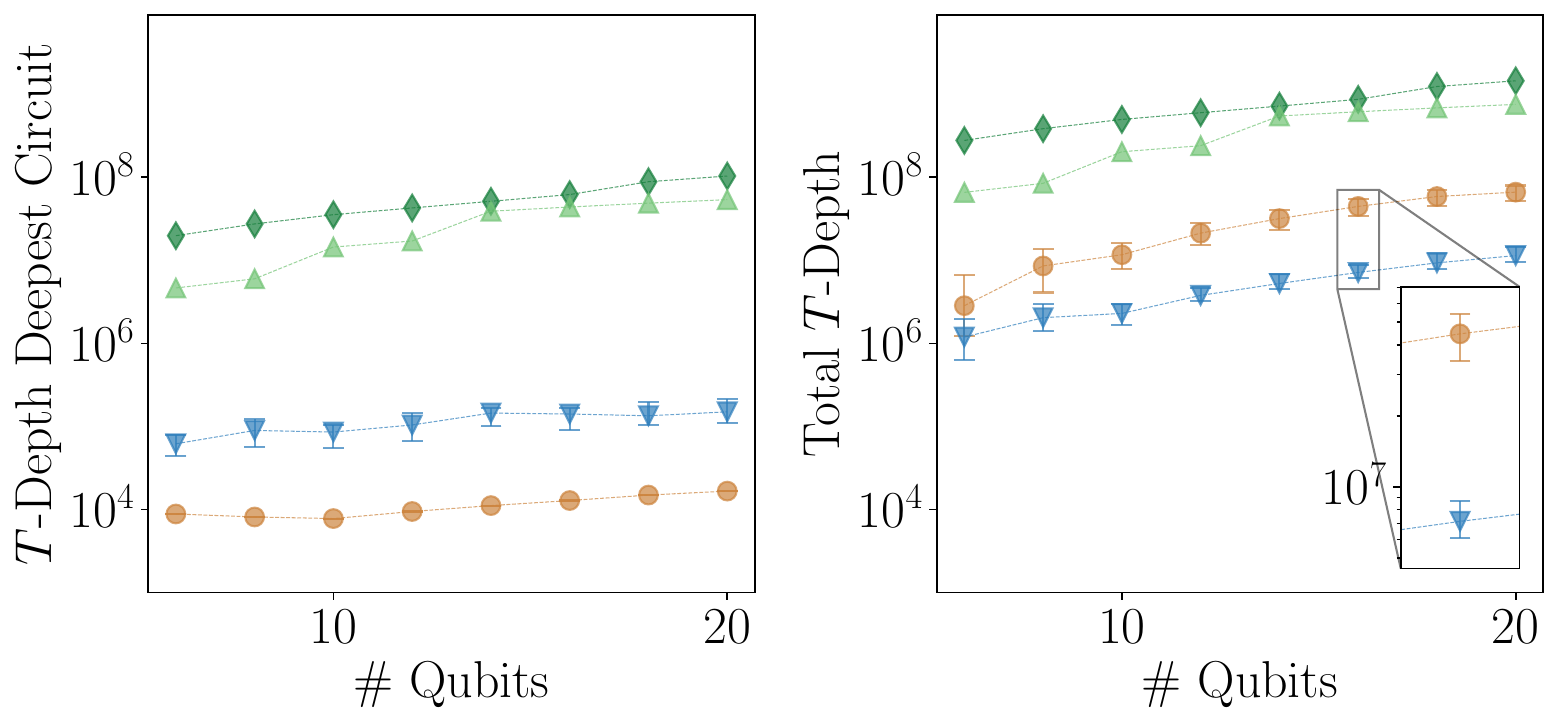}
    \caption{\textbf{$\boldsymbol{T}$-count (upper panels) and $\boldsymbol{T}$-depth (lower panels) to obtain the GSE of cyclopentadiene within $\boldsymbol{\epsilon = 1.6 \times 10^{-3}}$\,Ha.} The left panels show $T$-count (upper left panel) and $T$-depth (lower left panel) of the deepest circuit, i.e., the highest number of $T$-gates that are executed within one shot. The right panels display the total $T$-count (upper right panel) and total $T$-depth (lower right panel) which is the $T$-count / $T$-depth  multiplied with the total number of shots. We plot median values of 100 repetitions, with error bars representing the 68\,\% range of the restarts. Here, $\Ft=0.8$, $\Ns=10$, and $\tau=0.4$. The inset shows a zoomed-in view of the $T$-complexity of \ac{SQD-AA} and \ac{SQD} for 16 qubits.}
    \label{fig:hqs_Tgates}
\end{figure}

The $T$-counts and $T$-depths for \ac{SQD-AA}, \ac{SQD}, and \ac{iQPE} versus the number of qubits $n$ for cyclopentadiene are shown in \Fig \ \ref{fig:hqs_Tgates}. Note that the parameters for SQD-AA (i.e., $\Ns$, $\Ft$, and $\tau$) are discussed in Appendix \ref{subsec:params_sqdaa}. We show the total $T$-count ($T$-count $\times$ shots) and the total $T$-depth ($T$-depth $\times$ shots) in the right panels of \Fig \ \ref{fig:hqs_Tgates}, and the $T$-count and $T$-depth of the deepest circuit, i.e., the largest number of $T$-gates that are executed in one shot, in the left panels of \mbox{\Fig \ \ref{fig:hqs_Tgates}}. Here, the $T$-depth is the minimal number of sequential layers of $T$-gates, when $T$-gates acting on different qubits can be executed in parallel. 

For all system sizes, we observe the lowest total $T$-count for \ac{SQD-AA} (upper right panel of \Fig \ \ref{fig:hqs_Tgates}). 
The total $T$-count for \ac{SQD} is up to $\sim 6$ times higher, where the gap is mostly increasing with system size. This improvement is caused by a reduction in the sample complexity by a factor of $\sim 40$.
Moreover, as we show in the Appendix, \Fig \ \ref{fig:gain_comp_hqs}, we could obtain a reduction in the total query complexity of a factor of $\sim 10$ when sampling directly from the exact ground state. This is caused by the fact that probabilities of required bitstrings are higher in the adiabatically prepared state. Therefore, fewer shots $\Nsdir$ are required to measure all important bitstrings, and the overhead due to $\Ns$ for \ac{SQD-AA} carries more weight. The $T$-count for \ac{iQPE} with Qubitization is slightly smaller than the $T$-count for \ac{iQPE} with Trotterization, and both are roughly two orders of magnitude larger than for \ac{SQD}-methods for the considered system sizes.

In the upper left panel of \Fig \ \ref{fig:hqs_Tgates} we observe that the deepest circuit for \ac{SQD-AA} is roughly one order of magnitude deeper than the one for \ac{SQD}. In contrast, the circuits for \ac{iQPE} are both several orders of magnitude deeper, and the gap is increasing with system size.
In the lower panels of \Fig \ \ref{fig:hqs_Tgates} we show the same plots for the $T$-depth. 
For \ac{iQPE} with Trotterization and for the \ac{ASP} circuits, the $T$-depth amounts to $\sim 70\,\%$ of the $T$-count, since only few rotations act on disjoint qubits and can be executed in parallel (see Appendices~\ref{subsec:state_prep} and \ref{subsec:iqpe_trotter}). For \ac{iQPE} with Qubitization, the $T$-depth is roughly half of the $T$-count.  Hence, overall, we see similar trends as for the $T$-count.

When we consider early fault-tolerant quantum computing, only a limited number of logical $T$-gates can be executed with sufficiently low logical error rates. As $T$-count and $T$-depth of the deepest circuit are several orders of magnitude deeper for \ac{iQPE} compared to \ac{SQD}-methods, this suggests a regime where \ac{SQD-AA} can be executed while errors are too high for executing iQPE. To make this more concrete, a circuit comprising $G_C$ logical operations fails with a probability of at most $G_C\, p_L$, where $p_L$ denotes the logical error rate per operation, such that a circuit error rate below $p_C$ requires $p_L \leq p_C/G_C$~\cite{katabarwa_early_2024}. Using the $T$-count of the deepest circuit as a proxy for $G_C$, the deepest \ac{SQD-AA} circuits for cyclopentadiene ($\sim 10^5$ $T$-gates) can already be executed with logical error rates of $p_L \sim 10^{-6}$, as envisioned for the next generation of error-corrected `megaquop' machines~\cite{preskill_beyond_2025}, whereas the deepest \ac{iQPE} circuits ($\sim 10^7$--$10^8$ $T$-gates) require logical error rates that are two to three orders of magnitude lower. This is a crucial finding: sample-based diagonalization methods have so far mainly been considered in \ac{NISQ} settings, and our results strongly suggest an early fault-tolerant regime where these methods are feasible, while \ac{iQPE} can not be conducted confidently.

\begin{figure}
    \centering
    \includegraphics[width=\columnwidth]{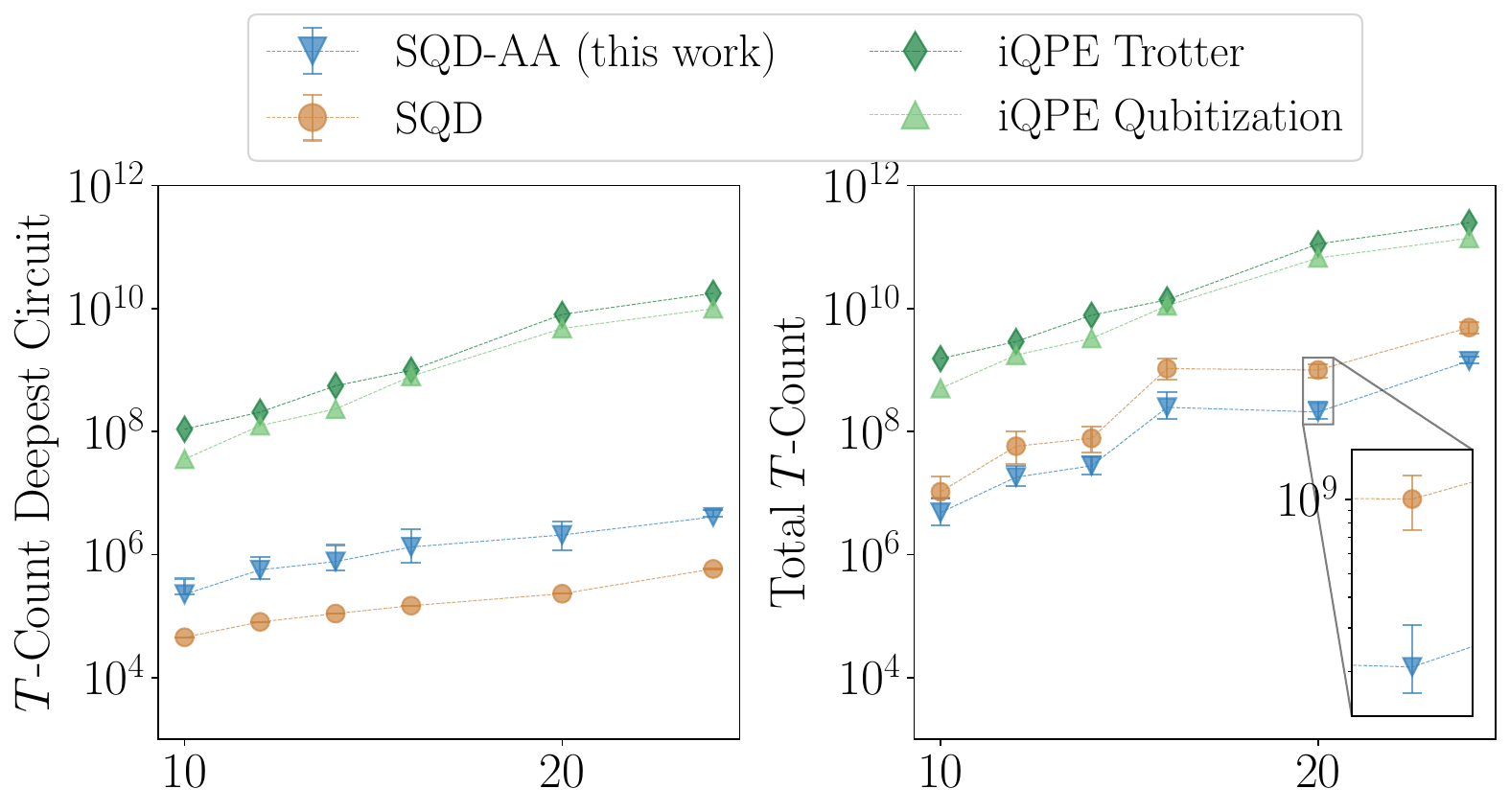}
    \includegraphics[width=\columnwidth]{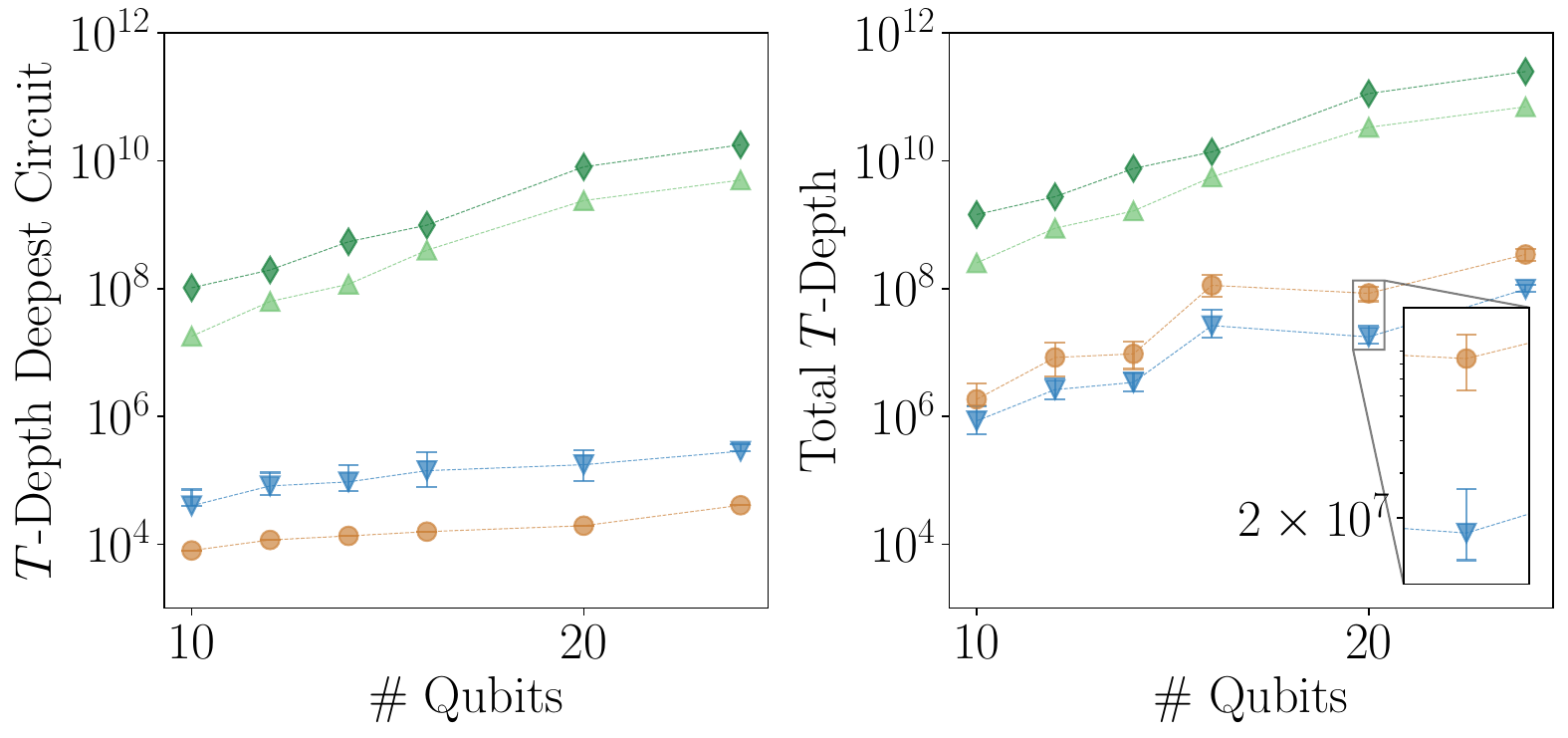}
    \caption{\textbf{$\boldsymbol{T}$-count (upper panels) and $\boldsymbol{T}$-depth (lower panels) to obtain the GSE of Cr$_{\boldsymbol{2}}$ within $\boldsymbol{\epsilon = 1.6 \times 10^{-3}}$\,Ha.} The left panels show $T$-count (upper left panel) and $T$-depth (lower left panel) of the deepest circuit, i.e., the highest number of $T$-gates that are executed within one shot. The right panels display the total $T$-count (upper right panel) and total $T$-depth (lower right panel) which is the $T$-count / $T$-depth multiplied with the total number of shots. We plot median values of 100 repetitions, with error bars representing the 68\,\% range of the restarts.  Here, $\Ns=10$ for \mbox{$n\leq20$} and $\Ns=100$ for $n>20$, $\Ft=0.8$, and $\tau=0.4$. The inset shows a zoomed-in view of the $T$-complexity for \ac{SQD-AA} and \ac{SQD} for 20 qubits.}
    \label{fig:cr2}
\end{figure}

To test if our findings are more broadly applicable, we present the same results for other molecules. First, we compare the different methods for the chromium dimer Cr$_2$ which is known to be challenging for classical methods \cite{larsson_chromium_2022, kanno_quantum-selected_2023}. We construct Hamiltonians for Cr$_2$ in different active spaces and employ the Jordan-Wigner mapping to encode the Hamiltonians in a qubit basis. (For details see Appendix \ref{subsec:molecules}.)  To implement $\Uprep$, we choose a classically optimized \ac{UCJ} ansatz \cite{motta_bridging_2023, robledo-moreno_chemistry_2025, shivpuje_sample-based_2025} that is elaborated in Appendix \ref{subsec:state_prep}. We select the smallest number of layers where all important bitstrings have non-vanishing probabilities. $T$-count and $T$-depth for \ac{SQD-AA}, \ac{SQD}, and \ac{iQPE} are plotted in \Fig \ \ref{fig:cr2}.

The total $T$-count, shown in the upper right panel of \Fig \ \ref{fig:cr2}, is highest for both \ac{iQPE} methods, Trotter and Qubitization, where we observe a similar $T$-count for both. For \ac{SQD-AA}, we obtain the lowest total $T$-count of all methods, especially for a higher number of qubits. Moreover, we get a reduction in the total $T$-count (i.e., in the total runtime) up to a factor of $\sim 5$ compared to \ac{SQD}. This runtime reduction corresponds to a reduction in the total number of shots by a factor of $\sim 35$. The $T$-count of the deepest circuit (upper left panel of \Fig \ \ref{fig:cr2}) is up to 3 orders of magnitude shallower for \ac{SQD-AA} compared to \ac{iQPE}. Compared to cyclopentadiene, the differences are even more pronounced. Moreover, the gap is increasing with increasing number of qubits, suggesting that also for this system there is an area where \ac{SQD-AA} can be executed while circuits are too deep to run for \ac{iQPE}.
When considering the $T$-depth (lower panels of \mbox{\Fig \ \ref{fig:cr2}}), we can see that the $T$-depth is significantly smaller than the $T$-count for \ac{SQD} and \ac{SQD-AA}. This is because many $T$-gates can be parallelized in the \ac{UCJ} ansatz. Therefore, the gap to \ac{iQPE} methods is even larger in this case and we observe improvements up to 4 orders of magnitude in the $T$-depth of the deepest circuit for \ac{SQD-AA}.

Yet, the subspaces for Cr$_2$ with the considered system sizes are relatively small ($m<50$) and hence, the overhead caused by $\Ns$ is rather large. This suggests that for systems where more shots are required for a target energy error, i.e., when the distribution is more rapidly decaying, a higher reduction in the $T$-count will be possible. To test this, we show results for molecules with larger subspace dimensions, H$_2$O and Mo$_2$, in Appendix~\ref{subsec:dif_sys}.

Here, we observe similar trends, i.e., the deepest circuit for \ac{iQPE} is several orders of magnitude deeper than for \ac{SQD}-methods, where the gap is increasing with system size. Moreover, we observe the lowest total $T$-count and $T$-depth for \ac{SQD-AA} and a reduction of up to a factor of  $\sim 10$ compared to \ac{SQD}, corresponding to a runtime reduction of one order of magnitude. In this case, the sample complexity is reduced by up to a factor of $\sim 65$. However, for H$_2$O we already observe a lower $T$-count for \ac{iQPE} with qubitization than for \ac{SQD}-methods for 24 qubits. Moreover, for Mo$_2$ the gap in the total $T$-count (and $T$-depth) between \ac{iQPE} and \ac{SQD} methods also decreases with system size. Therefore, we expect that \ac{iQPE} is often more efficient for larger systems. However, even if the total $T$-complexity of \ac{iQPE} might be smaller, the $T$-count of the deepest circuit that is executed within one shot is often several orders of magnitude deeper than for \ac{SQD}-methods. Moreover,
for all considered systems, the gap in the $T$-count of the deepest circuit is increasing with system size. Hence, we expect that a regime exists, where \ac{SQD}-methods are favorable.

\section{Discussion\label{sec:discussion}}
Within this work, we demonstrated that the sample complexity in \ac{SQD} can be substantially reduced via amplitude amplification. To achieve this, we introduced an algorithm, \ac{SQD-AA}, to reduce the probabilities of dominant bitstrings sequentially. This can significantly reduce the required measurement shots, albeit at the cost of deeper circuits. We showed that, for an exponentially decaying distribution, a quadratic advantage in the total query complexity over \ac{SQD} can be obtained for sufficiently large subspaces.
Note that in the context of Grover's algorithm, it has been claimed that a quadratic speedup is insufficient for potential quantum advantage in the foreseeable future \cite{hoefler_disentangling_2023}. This reasoning, however, does not apply to our setting, since one would only use a quantum computer in our case if the preparation of the initial state is already classically hard or intractable. Therefore, the quadratic speedup of \ac{SQD-AA} can still reduce the total runtime compared to \ac{SQD}, potentially from a year to a few days.

We further confirmed our findings in applications to real quantum chemical systems, where we assumed an early fault-tolerant scenario and compared the $T$-complexity to reach chemical accuracy within the active space for \ac{SQD-AA}, \ac{SQD}, and \ac{iQPE}. For almost all considered systems, we obtained the lowest $T$-complexity for \ac{SQD-AA}.
Importantly, the highest number of $T$-gates in a single circuit is several orders of magnitude larger for \ac{iQPE} than for \ac{SQD}-methods, suggesting that sample-based diagonalization methods will be viable in an early fault-tolerant setting.
Compared to \ac{SQD}, \ac{SQD-AA} improves the total $T$-complexity up to a factor of 10 for these example molecules.
Still, since we observe a reduction in the total query complexity of more than a factor of 100 for the algebraically and the exponentially decaying distribution, we expect that a higher reduction in the total runtime is possible for more rapidly decaying distributions.
Hence, we expect that in an early fault-tolerant scenario, when it is only possible to run circuits with a limited logical $T$-count at sufficiently low logical error rates, \ac{SQD-AA} can reliably be executed with orders of magnitude lower runtime than \ac{SQD}, while errors are still too high for running \ac{iQPE}.
\section*{Acknowledgements}
We thank Javier Robledo-Moreno and Etienne Granet for their valuable feedback on this manuscript.
This work is part of the Munich Quantum Valley, which is supported by the Bavarian state government with funds from the Hightech Agenda Bayern Plus.

\section*{Data Availability}
Data and code to reproduce the results of this work are available upon reasonable request.

\section*{Author Contributions}
N.S.: Conceptualization, Data curation, Formal analysis, Investigation, Methodology, Software, Validation, Visualization, Writing -- original draft, Writing -- review and editing. L.N.: Conceptualization, Methodology, Software, Writing -- original draft, Writing -- review and editing. M.J.H.: Conceptualization, Funding acquisition, Project administration, Supervision, Writing -- original draft, Writing -- review and editing.

\section*{Competing Interests}
M.J.H. is a shareholder of Quint Computing GmbH.

\begin{acronym}
\acro{AA}{amplitude amplification}
\acro{ASP}{adiabatic state preparation}

\acro{CASCI}{complete active space configuration interaction}
\acro{CCSD}{coupled cluster, singles and doubles}
\acro{UCCD}{unitary coupled cluster, doubles}
\acro{UCCSD}{unitary coupled cluster, singles and doubles}

\acro{full CI}{full configuration interaction}
\acro{FASQ}{fault-tolerant application-scale quantum}

\acro{GSE}{ground-state energy}
\acroplural{GSE}[GSEs]{ground-state energies}

\acro{HF}{Hartree-Fock}
\acro{HPC}{high-performance computing}

\acro{iQPE}{iterative quantum phase estimation}
\acro{ITE}{imaginary time evolution}

\acro{JW}{Jordan-Wigner}

\acro{LCU}{linear combination of unitaries}
\acro{LUCJ}{local unitary cluster Jastrow}

\acro{MOs}{molecular orbitals}
\acro{NISQ}{noisy intermediate-scale quantum}

\acro{QROM}{quantum read-only memory}
\acro{QPE}{quantum phase estimation}
\acro{QSCI}{quantum-selected configuration interaction}

\acro{RPA}{random phase approximation }
\acro{RUS}{repeat-until-success}

\acro{SQD}{sample-based quantum diagonalization}
\acro{SQD-AA}{sample-based quantum diagonalization with amplitude amplification}
\acro{SK}{Solovay-Kitaev}

\acro{UCJ}{unitary cluster Jastrow}

\acro{VQEs}{variational quantum eigensolvers}
\end{acronym}

\appendix
\setcounter{figure}{0}
\setcounter{table}{0}
\renewcommand{\thefigure}{A\arabic{figure}}
\renewcommand{\thetable}{A\arabic{table}}
\makeatletter
\renewcommand{\theHfigure}{appendix.\thefigure}
\renewcommand{\theHtable}{appendix.\thetable}
\makeatother
\section{Analysis of \ac{SQD-AA}\label{sec:appA}}
In this section, we evaluate \ac{SQD-AA} and \ac{SQD} for different distributions, followed by a comparison of \ac{SQD-AA} using \ac{AA} and fixed-point \ac{AA}.

\subsection{Analytic Comparison of \ac{SQD-AA} and \ac{SQD} for different Distributions}\label{subsec:sample_comp_sqdaa}

Here, we provide a detailed derivation of the results of Section \ref{subsec:dif_model_dists}, comparing \ac{SQD-AA} and \ac{SQD} for different model distributions. For all cases, we estimate the total query complexity to obtain the $m$ most probable bitstrings as a measure for the total runtime. First, we consider an exponentially decaying state (a), where we expect AA iterations to be more efficient than direct sampling. Next, we consider a distribution that follows a step-function (b), where all important bitstrings have the same probability, while all other probabilities are zero. This state would be the ideal initial state for bare \ac{SQD}. We conclude the analysis with an algebraically decaying state (c), which resembles a combination of the previous cases, i.e., probabilities decay strongly first, while the distribution becomes more flat at larger subspace dimensions, which is also often a feature of ground states of real molecules.\\

\textit{a) Exponentially decaying distribution:}
As stated previously, we consider the total query complexity to measure the $m$ most probable bitstrings. Note that for the exponentially decaying state we reduce the probabilities of all measured bitstrings, i.e., $m^* = m$, as the distribution is always sufficiently decaying.
We sequentially want to reduce the probabilities of these bitstrings, i.e., we want to rotate the initial state
\begin{align}
    \ket{\Psi_{\mathrm{exp}}} = \frac{1}{\sqrt{\sum_{l=0}^{N-1} e^{-\alpha l}}} \sum_{l=0}^{N-1} e^{-\alpha l/2} \ket{l}
\end{align}
to the target states
\begin{align}
    \ket{\phi_{\mathrm{t},k}} = \frac{1}{\sqrt{\sum_{l=k+1}^{N-1} e^{-\alpha l}}} \sum_{l=k+1}^{N-1} e^{-\alpha l/2} \ket{l}
\end{align}
for $k=0,1,\ldots,m-2$. Here, $N=2^n$ where $n$ is the number of qubits. The number of steps to reduce the probabilities of bitstrings $\{z_i\}_{i=0}^k$ is estimated as
\begin{align}
    s_{k+1} &= \left \lfloor \frac{\pi}{4 \theta_k} \right \rfloor \approx \left \lfloor \frac{\pi}{4 \sin(\theta_k)} \right \rfloor \nonumber  \\ &= \left \lfloor \frac{\pi}{4 \langle \Psi_{\mathrm{exp}} | \phi_{\text{t} ,k} \rangle} \right \rfloor \approx \left \lfloor \frac{\pi \sqrt{e^{\alpha (k+1)}}}{4} \right \rfloor,
    \label{eq_app:steps_exp_dec}
\end{align}
where we use $\theta_k \ll 1$ and evaluate $\sin(\theta_k)$ as
\begin{align}
    \sin(\theta_k) &= \langle \Psi_{\mathrm{exp}} | \phi_{\text{t},k} \rangle \approx  \frac{\sqrt{1-e^{-\alpha}}}{\sqrt{\sum_{l=k+1}^{N-1} e^{-\alpha l}}} \sum_{l=k+1}^{N-1} e^{-\alpha l}  \nonumber \\ 
    &= \sqrt{1-e^{-\alpha}} \sqrt{\sum_{l=k+1}^{N-1} e^{-\alpha l}} 
    \approx \sqrt{e^{-\alpha (k+1)}},
\end{align}
making use of the geometric series
\begin{align}
    \sum_{l=k}^{N-1} e^{-\alpha l} = \frac{e^{-\alpha k}- e^{-\alpha N}}{1 - e^{-\alpha}} \approx \frac{e^{-\alpha k}}{1 - e^{-\alpha }}.
\end{align}
Note that the steps of iteration $k=0$ are zero, $s_0=0$, i.e., only $\Uprep$ is applied in the first iteration.

Denoting by $Q_{\mathrm{tot,AA}}^\mathrm{SQD-AA}$ the total query complexity, i.e., the total number of times $\Uprep$ is applied during AA iterations, we have
\begin{align}
    Q_{\mathrm{tot,AA}}^\mathrm{SQD-AA} &= \Ns \cdot \sum_{k=0}^{m-1} Q_k  \nonumber \\   
    &=  \Ns \cdot \sum_{k=0}^{m-1} (2 s_k +1)  \nonumber \\
    &\approx \Ns \cdot \left(m + \frac{\pi}{2} \sum_{k=0}^{m-1} \sqrt{e^{\alpha k}} \right)  \nonumber \\ 
    &= \Ns \cdot \left(m +  \frac{\pi}{2} \cdot \frac{\sqrt{e^{\alpha m}} -1}{e^{\alpha/2} -1} \right),
    \label{eq_app:s_exp_sqdaa}
\end{align}
where we approximate $2\lfloor \pi \sqrt{e^{\alpha k}} / 4 \rfloor \leq \pi \sqrt{e^{\alpha k}} / 2$. Note that for the exponentially decaying state the probability of the most probable bitstring is the same for each $\ket{\Psi_k}$ (assuming $m \ll N$ and an ideal reduction of all $\{p_i\}_{i=0}^{k-1}$ to zero). Thus, as already mentioned, the distribution is always sufficiently decaying such that AA is more efficient than direct sampling, and $\Qtot^\mathrm{SQD-AA}= Q_{\mathrm{tot,AA}}^\mathrm{SQD-AA}$. 
Yet, it is worth noting that multiple unique bitstrings could be discovered within one iteration, which is not considered in this analysis. This enhances the quadratic improvement slightly, as can be seen in the numerical simulations (cf. \Fig \ \ref{fig:expalgcomp}).

For bare \ac{SQD}, we choose $\Nsdir$ such that the probability $p_{\mathrm{fail}}$ of not seeing one of the first $m$ bitstrings is upper bounded by
\begin{align}
    p_{\mathrm{fail}} &:= \sum_{k=0}^{m-1} (1-p_k)^{\Nsdir} \leq \sum_{k=0}^{m-1} (1-p_{m-1})^{\Nsdir}  \nonumber \\ 
    &\leq m \cdot e^{-\Nsdir\, p_{m-1}},
    \label{eq_app:pfail}
\end{align}
where we use that the bitstrings are ordered by decreasing probability, $p_{k+1} \leq p_k$. As we apply $\Uprep$ once for each shot, the total query complexity for \ac{SQD} equals the shot count $\Nsdir$ and we have
\begin{align}
    \Qtot^\mathrm{SQD} &= \Nsdir \geq \frac{1}{p_{m-1}} \ln{\frac{m}{p_{\mathrm{fail}}}}  \nonumber \\
    & = \frac{e^{\alpha (m-1)}}{1 -e^{-\alpha}} \ln{\frac{m}{p_{\mathrm{fail}}}}.
    \label{eq_app:s_exp_sqd}
\end{align}
Thus, for an exponentially decaying distribution, the total number of times that $\Uprep$ is applied scales as $\sqrt{e^{\alpha m}}$ for \ac{SQD-AA}, while it increases $\propto e^{\alpha m}$ for \ac{SQD}. Hence, we obtain a quadratic advantage for \ac{SQD-AA} for sufficiently large $m$.\\

\textit{b) Step function-like distribution:}
The next distribution we analyze is a distribution that resembles a step function where all $m$ important bitstrings $\{z_i \}_{i=0}^{m-1}$ have probability $p_i = 1/m$, while $p_i = 0$ for $i \geq m$. That is, we rotate the state
\begin{align}
    \ket{\Psi_{\mathrm{step}}} = \sum_{l=0}^{m-1} \frac{1}{\sqrt{m}} \ket{l}
\end{align}
to the target states
\begin{align}
    \ket{\phi_{\mathrm{t},k}} = \frac{1}{\sqrt{\sum_{l=k+1}^{m-1} \frac{1}{m}}} \sum_{l=k+1}^{m-1} \frac{1}{\sqrt{m}} \ket{l}
\end{align}
for $k=0,1, \ldots,m-2$. The number of steps to reduce the probabilities of bitstrings $\{z_i\}_{i=0}^{k}$ is given by
\begin{align}
    s_{k+1} &= \left \lfloor \frac{\pi}{4 \theta_k} \right \rfloor = \left \lfloor \frac{\pi}{4 \arcsin(\langle \Psi_{\mathrm{step}} | \phi_{\text{t} ,k} \rangle)} \right \rfloor  \nonumber \\ 
    &= \left \lfloor \frac{\pi}{4 \arcsin\left(\frac{\sqrt{m-k-1}}{\sqrt{m}} \right)} \right \rfloor,
    \label{eq_app:steps_stepfct}
\end{align}
where $\langle \Psi_{\mathrm{step}} | \phi_{\text{t} ,k} \rangle$ is evaluated as
\begin{align}
    \langle \Psi_{\mathrm{step}} | \phi_{\text{t} ,k} \rangle = \sqrt{ \sum_{l=k+1}^{m-1} \frac{1}{m}} = \sqrt{\frac{m-k-1}{m}}.
\end{align}
Note that because $\ket{\Psi_{\mathrm{step}}}$ and $\ket{\phi_{\text{t} ,k}}$ have large overlap when considering a distribution following a step-function, the angles $\theta_k$ are not small enough to approximate $\theta_k$ as $\sin(\theta_k)$ in this case.
Using $x \leq \arcsin({x})$ for $x \in [0,1]$, we estimate $Q_\mathrm{tot,AA}^\mathrm{SQD-AA}$ as
\begin{align}
    Q_\mathrm{tot,AA}^\mathrm{SQD-AA} &=  \Ns \cdot \sum_{k=0}^{m-1} (2 s_k +1)  \nonumber \\ &\approx \Ns \cdot \left(m + \frac{\pi}{2} \sum_{k=0}^{m-1} \frac{1}{\arcsin \left(\frac{\sqrt{m-k}}{\sqrt{m}}\right)} \right)  \nonumber \\ &\leq \Ns \cdot \left(m + \frac{\pi}{2} \sum_{k=0}^{m-1} \frac{\sqrt{m}}{\sqrt{m-k}}\right)  \nonumber \\
    &\leq \Ns \cdot m (1 + \pi).
\end{align}
In the last line we use
\begin{align}
    \sum_{k=0}^{m-1} \frac{\sqrt{m}}{\sqrt{m-k}}= \sqrt{m} \sum_{j=1}^{m} \frac{1}{\sqrt{j}} \leq 2 m,
\end{align}
where we reverse the order of the sum and use the integral test to upper bound the monotonically decreasing sum,
\begin{align}
    \sum_{j=1}^{m} \frac{1}{\sqrt{j}} \leq 1 + \int_1^{m} \frac{1}{\sqrt{x}} dx = 2\sqrt{m} -1.
\end{align}

For SQD, we again estimate $\Nsdir$ via \Eq \ \eqref{eq_app:pfail}, and thus obtain the total query complexity
\begin{align}
    \Qtot^\mathrm{SQD} = m \ln{\frac{m}{p_{\mathrm{fail}}}}.
\end{align}
For reasonable subspace sizes, $\Ns \cdot (1+\pi) > \ln(m/p_{\mathrm{fail}})$, which means that we cannot obtain an advantage with SQD-AA. This suggests that for states that resemble a mixture of the two considered distributions, \ac{AA} should only be applied where the distribution decays sufficiently, followed by direct measurements of the remaining basis states. A distribution with this property is the algebraically decaying distribution, where $p_l \propto (l+1)^{-\gamma}$.\\

\textit{c) Algebraically decaying distribution:}
Here, probabilities decay strongly first, while the distribution flattens with increasing subspace dimension $m$, which is also often a feature of the ground states of quantum chemical Hamiltonians (cf. \Fig \ \ref{fig:gain_comp_hqs}).

As for the other states, we rotate the state
\begin{align}
    \ket{\Psi_{\mathrm{alg}}} = \frac{1}{\sqrt{\sum_{l=0}^{N-1} (l+1)^{-\gamma}}} \sum_{l=0}^{N-1} (l+1)^{-\gamma/2} \ket{l}
\end{align}
to the target states
\begin{align}
    \ket{\phi_{\mathrm{t},k}} = \frac{1}{\sqrt{\sum_{l=k+1}^{N-1} (l+1)^{-\gamma}}} \sum_{l=k+1}^{N-1} (l+1)^{-\gamma/2} \ket{l}
\end{align}
for $k=0,1,\ldots,m^*-2$, where $m^* \ll m$. The number of steps to reduce the probabilities of bitstrings $\{z_i\}_{i=0}^k$ is estimated as
\begin{align}
    s_{k+1} &= \left \lfloor \frac{\pi}{4 \theta_k} \right \rfloor \approx \left \lfloor \frac{\pi}{4 \langle \Psi_{\mathrm{alg}} | \phi_{\text{t} ,k} \rangle} \right \rfloor  \nonumber \\
    &\approx \left \lfloor \frac{\pi \sqrt{\sum_{l=0}^{N-1} (1+l)^{-\gamma} }}{4 \sqrt{\sum_{l={k+1}}^{N-1} (1+l)^{-\gamma}}} \right \rfloor
    \approx \left \lfloor \frac{\pi \sqrt{\zeta (\gamma)}}{4 \sqrt{\zeta (\gamma) - H_{k+1}(\gamma)}} \right \rfloor
    \label{eq_app:steps_alg_dec}
\end{align}
where we approximate the sum $\sum_{l=0}^{N-1} (l+1)^{-\gamma}$ via the Riemann zeta function $\zeta(\gamma)$ and define the $k$th harmonic number of order $\gamma$, $H_k(\gamma) = \sum_{l=0}^{k-1} (l+1)^{-\gamma}$. Next, we determine the total query complexity to obtain the $m$ most probable bitstrings. For the $m^*$ AA iterations, we get
\begin{align}
    Q_\mathrm{tot,AA}^\mathrm{SQD-AA} &= \Ns \sum_{k=0}^{m^*-1} Q_k  \nonumber \\
    &\approx \Ns \left(m^* +  \frac{\pi}{2} \cdot \sum_{k=0}^{m^*-1} \frac{\sqrt{\zeta (\gamma)}}{\sqrt{\zeta (\gamma) - H_k(\gamma) }}\right).
    \label{eq:qtot_aa_algdec}
\end{align}
The additional term, sampling the state after $m^*$ iterations, $\ket{\Psi_{m^*-1}}$, until all $m$ bitstrings are measured, reads
\begin{align}
    Q_\mathrm{tot,dir}^\mathrm{SQD-AA} &= N_\mathrm{S}^{\mathrm{AA,dir}}  Q_{m^*-1}  \nonumber \\
    &\approx m^{\gamma} (\zeta(\gamma)-H_{m^*-1}(\gamma))  \ln{\frac{m-m^*}{p_{\mathrm{fail}}}} Q_{m^*-1}  \nonumber \\
    &\approx m^{\gamma} \frac{\pi}{2} \sqrt{\zeta(\gamma)} \sqrt{(\zeta(\gamma)-H_{m^*-1}(\gamma))}  \ln{\frac{m-m^*}{p_{\mathrm{fail}}}}
    \label{eq_app:qtot_dir_alg_sqdaa}
\end{align}
where the number of shots is estimated according to \Eq \ \eqref{eq:pfail}. Moreover, $\zeta(\gamma)-H_{m^*-1}(\gamma) \ll 1$, i.e., far fewer shots are required to measure $\ket{\Psi_{m^*-1}}$ compared to measuring $\GS$. The total query complexity is then $Q_\mathrm{tot}^\mathrm{SQD-AA} = Q_\mathrm{tot,AA}^\mathrm{SQD-AA} + Q_\mathrm{tot,dir}^\mathrm{SQD-AA}$. Here, we choose $m^*$ so that $Q_\mathrm{tot}^\mathrm{SQD-AA}$ is minimal. In the actual Algorithm \ref{algo:sqdaa}, this is incorporated via the convergence criterion $\tau$.

For bare \ac{SQD}, we obtain
\begin{align}
    Q_\mathrm{tot}^\mathrm{SQD} &= \Nsdir \geq \frac{1}{p_{m-1}} \ln{\frac{m}{p_{\mathrm{fail}}}}  \nonumber \\
    &= m^{\gamma} \zeta(\gamma)  \ln{\frac{m}{p_{\mathrm{fail}}}}.
    \label{eq_app:qtot_exp_alg_sqd}
\end{align}
As these expressions are more complex, we refer to \Fig \ \ref{fig:expalgcomp} for a detailed comparison of the derived total query complexities. In summary, the analysis shows that \ac{SQD-AA} is most efficient for strongly decaying states and the advantage increases with subspace dimension $m$.

\subsection{Comparison of AA and fixed-point AA}\label{subsec:aa_fpaa}
As mentioned in Section \ref{sec:aa}, over-rotations can be avoided using the fixed-point version of \ac{AA} at the cost of a larger optimal number of steps $s_{k+1}$. Here, we first give a brief introduction to fixed-point AA \cite{yoder_fixed-point_2014} and then compare SQD-AA using standard AA and the fixed-point version for various systems.\\

As in standard \ac{AA}, we rotate $\GS$ toward a target state $\ket{\phi_{\mathrm{t},k}}$ via a series of reflections. In contrast, however, we use generalized reflections,
\begin{align}
    S_{P_k} = \mathbb{I} - (1-e^{\mathrm{i}\beta})P_k
\end{align}
where
\begin{align}
    P_k = \sum_{z_i \in \mathcal{S}_k} |z_i\rangle \langle z_i|
\end{align}
and
\begin{align}
    S_{\GSwf} =  \mathbb{I} - (1-e^{-\mathrm{i}\alpha}) |\tilde{\Psi}_{\mathrm{GS}} \rangle \langle \tilde{\Psi}_{\mathrm{GS}}|.
\end{align}
The two reflections generate a rotation $A_{k+1}(\alpha, \beta) =-S_{\GSwf}(\alpha)S_{P_k}(\beta)$. The angles $\alpha$ and $\beta$ are determined via Chebyshev polynomials for $j=1,2, \ldots, s_{k+1}$,
\begin{align}
    \alpha_j &= \beta_{s_{k+1}-j+1} \nonumber \\ &= 2 \cot^{-1}\left(\tan\left(\frac{2\pi j}{L_{k+1}}\right) \sqrt{1-\gamma^2} \right),
\end{align}
with $L_{k+1} = 2s_{k+1}+1$ the number of queries to $\Uprep$ and $\Uprep^\dagger$, and $\gamma^{-1} = T_{1/L_{k+1}}(1/\delta)$. The operator $A_{k+1}$ is applied $s_{k+1}$ times for different angles
\begin{align}
    A_{k+1}^{s_{k+1}} = \prod_{j=1}^{s_{k+1}} A_{k+1}(\alpha_j, \beta_j)
\end{align} 
where 
\begin{align}
    L_{k+1} = 2s_{k+1}+1 \geq \frac{\ln(2/\delta)}{|\langle \phi_{\mathrm{t},k} \GS|}
    \label{eq_app:steps_fpaa}
\end{align}
guarantees that
\begin{align}
    |\langle \phi_{\mathrm{t},k}|&A_{k+1}^{s_{k+1}}\GS |^2&  \nonumber \\ &= 1 - \delta^2 T_{L_{k+1}} \left(T_{1/L_{k+1}}(1/\delta) \sqrt{1 - |\langle \phi_{\mathrm{t},k} \GS |^2} \right )^2&
     \nonumber \\ &\geq 1 - \delta^2&
     \label{eq_app:overlap_fpaa}
\end{align}
Here,
\begin{align}
    T_{L}(x) = \cos(L \arccos(x))
\end{align}
is a Chebyshev polynomial of the first kind. Thus, adapting the angles guarantees that the fidelity of the rotated and the target state is above $1-\delta^2$. Choosing $1-\delta^2 = 0$ recovers the original AA where all $\alpha_j = \beta_j = \pi$. The ideal number of steps $s_{k+1,\mathrm{id}}$ is, however, not given by \Eq \ \eqref{eq_app:steps_fpaa} but can be determined by evaluating \Eq \ \eqref{eq_app:overlap_fpaa} for different $s_{k+1}' \geq s_{k+1}$ and choosing the number of steps for which the fidelity with the target state is closest to the target fidelity $\Ft$, as for standard \ac{AA}.

\begin{figure*}[t]
    \centering

    \begin{minipage}{0.32\textwidth}
        \centering
        \includegraphics[height=5.4cm]{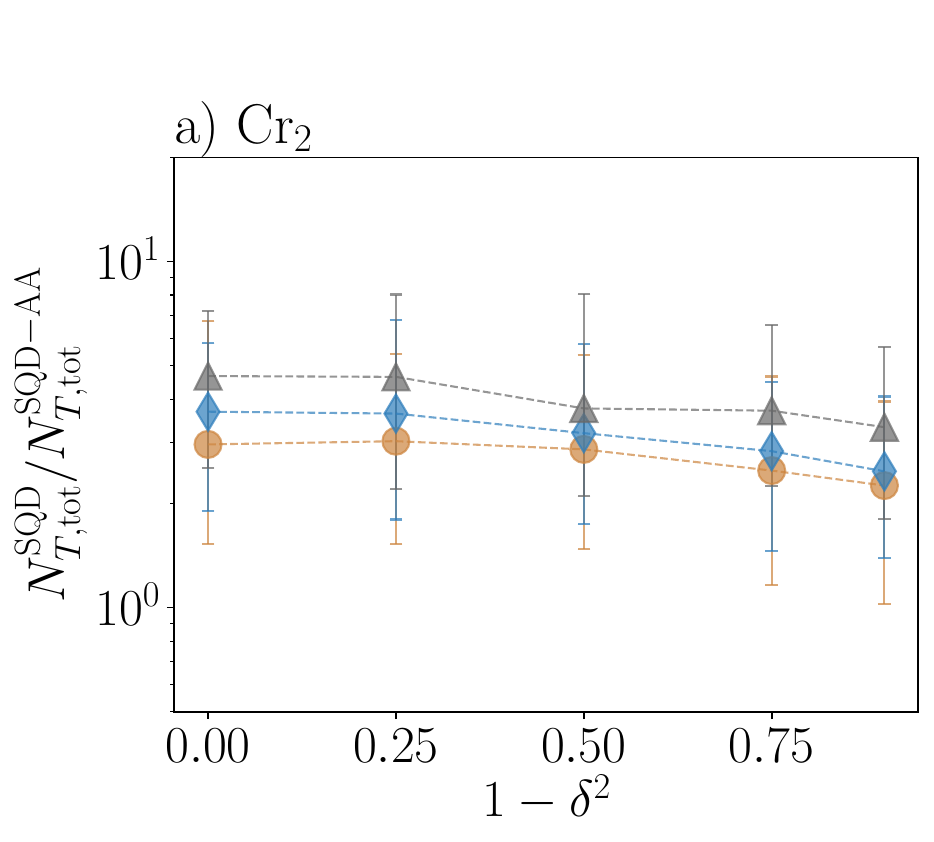}
    \end{minipage}\hfill
    \begin{minipage}{0.32\textwidth}
        \centering
        \includegraphics[height=5.4cm]{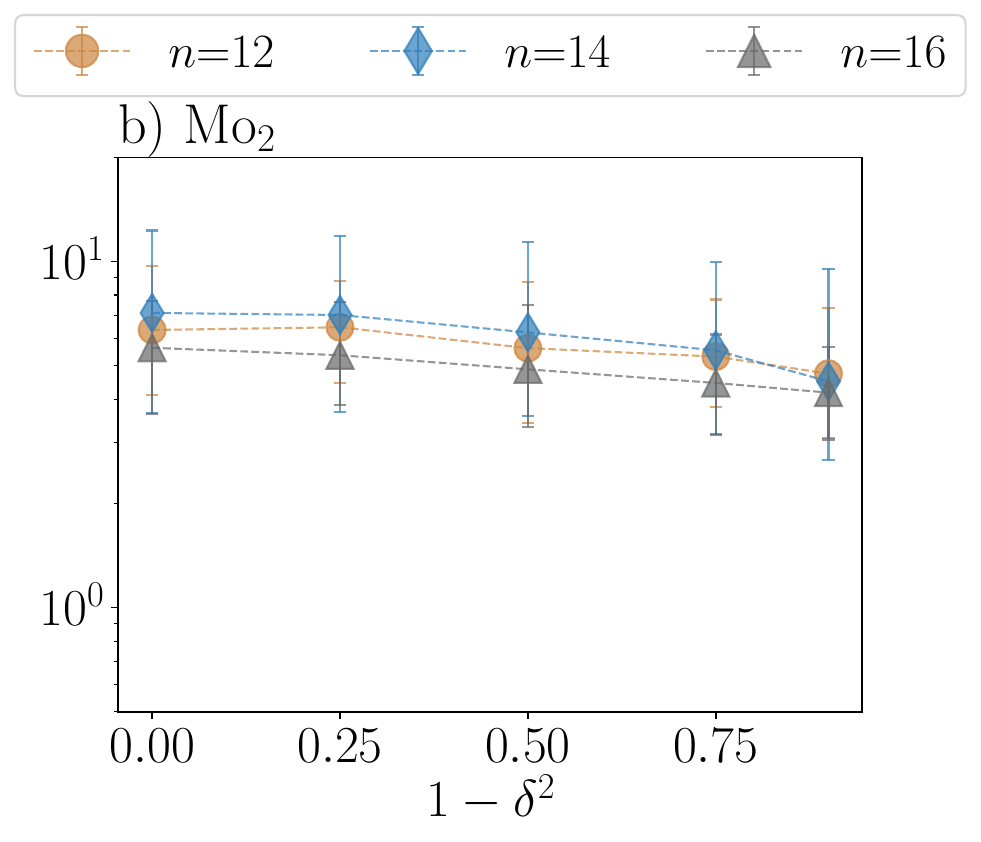}
    \end{minipage}\hfill
    \begin{minipage}{0.32\textwidth}
        \centering
        \includegraphics[height=5.4cm]{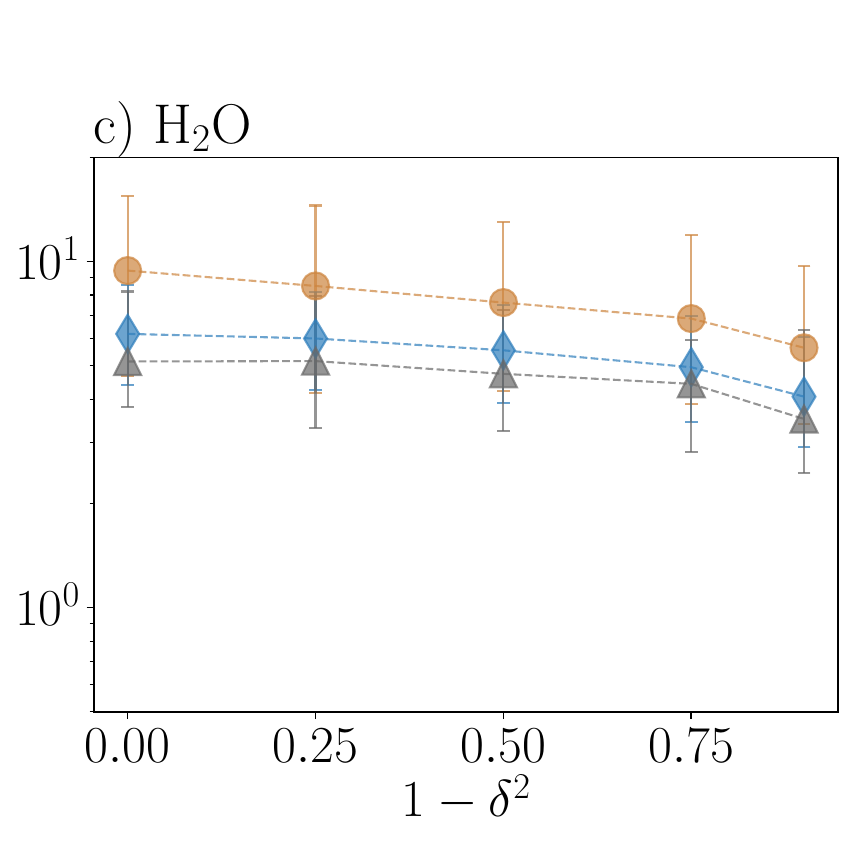}
    \end{minipage}

    \caption{\textbf{Median reduction in $T$-gates for different thresholds $\boldsymbol{1-\delta^2}$.} Results are shown for \textbf{a)} Cr$_2$, \textbf{b)} Mo$_2$ and \textbf{c)} H$_2$O using $\Ns=10$, $\mathcal{F}_{\mathrm{T}} = 0.8$, and $\tau=0.4$. For each system, we show results for 12, 14 and 16 qubits $n$. Error bars represent the 68\,\% range of 100 restarts.}
    \label{fig:fpaa_aa_dif_mols}
\end{figure*}

We now evaluate SQD-AA in its standard and fixed-point form for different molecules and system sizes using different values for $\delta$. In \Fig \ \ref{fig:fpaa_aa_dif_mols} we plot the reduction in total $T$-gates ($T$-count $\times$ shots) for Cr$_2$, Mo$_2$ and H$_2$O  against $1-\delta^2$.

For all systems and system sizes, we observe the highest reduction in the $T$-count for $1-\delta^2 = 0$ which corresponds to the original AA version where all $\alpha_j = \beta_j = \pi$. When increasing $1 - \delta^2$, the factor of reduction decreases and is lowest for $1 - \delta^2 = 0.9$. This is related to the larger ideal number of steps with increasing $1 - \delta^2$. Moreover, if the estimated number of steps is below $s_{k+1,\mathrm{id}}$, there is no advantage in using fixed-point AA. Therefore, we use the original AA throughout the paper; however, fixed-point AA can be used as well, and there might exist systems where this version is advantageous.

\section{Implementation of SQD-AA\label{sec:appB}}
In this section, we examine details on the implementation of SQD-AA. For that, we first discuss how Hamiltonians for the test molecules are constructed, followed by a description of the state preparation methods. Thereafter, we evaluate SQD-AA with different parameters (i.e., $\Ns$, $\mathcal{F}_T$, and $\tau$) and show results for different molecules.

\subsection{Quantum Chemical Methods}\label{subsec:molecules}
Within this paper, we use different formulations of the electronic structure Hamiltonian. For Cr$_2$, H$_2$O and Mo$_2$, we express the Hamiltonian in second-quantized form, whereas for cyclopentadiene an effective Hamiltonian is derived using \ac{RPA}. In the following, we detail the construction of these Hamiltonians.\\

\textit{a) Molecular Hamiltonian in second-quantization:} The electronic structure Hamiltonian in second-quantization is expressed in terms of fermionic creation ($a^{\dagger}$) and annihilation ($a$) operators and reads
\begin{align}
    	H_{\mathrm{el}} = \sum_{pq,\sigma} h_{pq}^{\sigma} a^\dagger_{p\sigma} a_{q\sigma} + \frac{1}{2} \sum_{pqrs, \sigma \sigma'} g_{pqrs}^{\sigma, \sigma'} a^\dagger_{p\sigma} a^\dagger_{q \sigma'} a_{s \sigma'} a_{r \sigma}.
	\label{app_eq:fermionic_h}
\end{align}
Here, $\{p,q,r,s\}$ label spatial orbitals, $\sigma \in \{\uparrow, \downarrow\}$ denotes the spin and $h_{pq}$ and $g_{pqrs}$ are one- and two-electron integrals, respectively \cite{miller_classical_1986}. To ensure antisymmetry of the wavefunction, the creation and annihilation operators obey anticommutation relations \cite{nielsen_fermionic_nodate},
\begin{align}
    \{ a_{p\sigma}, a_{q\tau}\} = 0, \quad \{ a_{p\sigma}^\dagger, a_{q\tau}^\dagger \} = 0, \quad \{ a_{p\sigma}, a_{q\tau}^\dagger \} = \delta_{pq}\delta_{\sigma\tau}.
\end{align}
The one- and two-body integrals are calculated classically using \ac{CASCI}. Within \ac{CASCI}, only a selected subset of spatial orbitals is treated exactly at \ac{full CI} level, while the remaining orbitals are handled using approximate methods such as \ac{HF} \cite{choe_multireference_2001}. To conduct the quantum chemical calculations, we use the \texttt{PySCF} package \cite{sun_recent_2020}. In Table \ref{tab:active_spaces}, we list the active spaces and basis sets that we use for different molecules. Moreover, geometries were taken from the NIST Computational Chemistry
Comparison and Benchmark Database (CCCBDB) \cite{nist_cccbdb_2022}.
\begin{table}[h]
\renewcommand{\arraystretch}{1.4}
\centering
\caption{Molecules, basis sets, and corresponding active spaces used in this work.}
\begin{adjustbox}{max width=\columnwidth}
\begin{tabular}{l l l}
\hline
Mol & Basis set & Active spaces ($n_{\mathrm{orb}}, n_{\mathrm{elec}}$) \\
\hline
Cr$_2$   & cc-pVDZ  \cite{balabanov_systematically_2005}   & (5,4), (6,6), (7,6), (8,8), (10,10), (12,12)  \\
H$_2$O  & cc-pVDZ \cite{dunning_gaussian_1989}   &   (5,4), (6,6), (7,6), (8,6), (10,10), (12,10) \\
Mo$_2$  & def2-SVP \cite{weigend_balanced_2005} & (5,6), (6,6), (7,8), (8,10), (10,10), (12,12) \\
\hline
\end{tabular}
\end{adjustbox}
\label{tab:active_spaces}
\end{table}

To transform the Hamiltonian to a qubit Hamiltonian, we employ the \ac{JW} transformation \cite{tranter_comparison_2018}. The mapped Hamiltonian is expressed as sum of Pauli strings, i.e., matrix elements during SQD-AA can be evaluated via parity rules.\\

\textit{b) Hamiltonian with random phase approximation (RPA):} The Hamiltonian for cyclopentadiene was obtained via \ac{RPA} by Refs.~\cite{shirazi_efficient_2024, hqs2026hqstage}. It consists of two active \ac{MOs} (described by four qubits) that are coupled to a bath. Additional qubits determine the number of environment orbitals that are taken into account. Increasing the number of qubits therefore increases the accuracy in the \ac{GSE}. The system is used in electron spectroscopy, where the singlet-triplet gap $\Delta_{S,T}$, i.e., here the gap between the ground and first-excited state is of interest. In Section \ref{subsec:benchmark_sqdaa} we show results obtaining the \ac{GSE} with SQD-AA. To obtain the spectral gap, however, the first-excited energy must be determined. For cyclopentadiene, two bitstrings are sufficient to determine the first-excited energy within an energy error of $10^{-15}$\,Ha using SQD. We assume that these two bitstrings can be efficiently obtained with classical methods and, hence, we use the quantum computer only for the \ac{GSE}.

\subsection{State Preparation}\label{subsec:state_prep}
For the electronic-structure Hamiltonian (a), we use a classically pre-optimized \ac{UCJ} ansatz to prepare an approximate ground state. In contrast, for the \ac{RPA} Hamiltonian (b), we apply adiabatic state preparation (\ac{ASP}), which is suitable here as this Hamiltonian contains only a low number of Pauli strings.\\

\textit{a) \ac{UCJ} ansatz:} The \ac{UCJ} ansatz $\ket{\Psi_{\mathrm{UCJ}}}$ is derived as Trotter approximation of a double-factorized form of the \ac{UCCD} ansatz \cite{motta_bridging_2023, evangelista_exact_2019, matsuzawa_jastrow-type_2020}
\begin{align}
    e^{T_2-T_2^{\dagger}} \ket{\Psi_{\mathrm{HF}}} &= e^{\mathrm{i}\sum_{\mu=1}^{L} e^{K_{\mu}} J_{\mu} e^{-K_{\mu}}}  \ket{\Psi_{\mathrm{HF}}} \label{eq_app:uccsd_t2_df} \nonumber \\ &\approx \prod_{\mu=1}^{L} e^{K_{\mu}} \, e^{\mathrm{i} J_{\mu}} \, e^{-K_{\mu}} \ket{\Psi_{\mathrm{HF}}} \nonumber \\
    &\equiv \ket{\Psi_{\mathrm{UCJ}}}.
\end{align}
Here, the double-excitation operator $T_2$ denotes
\begin{align}
    T_2 = \sum_{ijrs} t_{ijrs} \, a_r^{\dagger} a_{s}^{\dagger} a_j a_i
\end{align}
where $\{i,j\}$ index occupied and $\{r,s\}$ unoccupied \ac{MOs}. The ansatz consists of $L$ layers of orbital rotations $e^{K_{\mu}}$ and exponentials of diagonal Coulomb operators $J_{\mu}$ where
\begin{align}
    K_{\mu} = \sum_{pq, \sigma} K_{pq}^{\mu} a_{p\sigma}^{\dagger} a_{q\sigma}, \ \ \ J_{\mu} = \sum_{pq, \sigma \tau} J_{pq,\sigma \tau}^{\mu} n_{p\sigma} n_{q\tau}.
\end{align}
The indices $\{p,q\}$ describe spatial \ac{MOs}, $\{\sigma, \tau \}$ label spin polarizations, and $n = a^{\dagger} a$ is the number operator. We use the spin-balanced version where $J^{\uparrow \uparrow} = J^{\downarrow \downarrow}$ and $J^{\uparrow \downarrow} = J^{\downarrow \uparrow}$, i.e., each diagonal Coulomb operator is expressed by two symmetric matrices.

The $T_2$ operator can be efficiently obtained on a classical computer via \ac{CCSD} calculations. A subsequent double factorization of $T_2 - T_2^{\dagger}$ then provides the parameters $K_{pq}^{\mu}$ and $J_{pq,\sigma \tau}^{\mu}$ for the \ac{UCJ} ansatz (see \Eq\  \eqref{eq_app:uccsd_t2_df}). While $T_2$ only contains double excitations, single excitations may also be included as a final orbital rotation. Because the double factorization of $T_2$ is often approximately low-rank, truncating the number of layers $L$ can still produce accurate results while substantially reducing computational cost \cite{motta_low_2021}. In this work, we test different numbers of layers and choose the lowest $L$ such that all necessary bitstrings for a target energy error have non-vanishing probabilities. When using a lower number of layers, chemical accuracy in the active space might not be reached with \ac{SQD}, as some important bitstrings could have probabilities close to zero. The layers are listed in Table \ref{tab:ucj_layers}. We also remark that we do not use any locality constraints, as we assume all-to-all connectivity on the (early) fault-tolerant device.

\begin{table}[ht!]
\renewcommand{\arraystretch}{1.4}
\centering
\caption{Layers $L$ of the \ac{UCJ} ansatz for different molecules and qubit numbers $n$. We use the minimal $L$ such that all bitstrings for an energy error of $1.6 \times 10^{-3}$\,Ha have non-vanishing probabilities.}
\begin{tabular}{l rrrrrr}
\hline
 Mol & \multicolumn{6}{c}{\ac{UCJ} layers ($L, n$)} \\
\hline
Cr$_2$ \hspace{1.6cm}  &  (5,10)& (6,12)& (6,14)& (6,16)& (6,20)& (10,24)  \\
H$_2$O \hspace{1.6cm} &    (2,10)& (3,12)& (5,14)& (9,16)& (17,20)& (20,24)\\
Mo$_2$ \hspace{1.6cm} &  (4,10)& (5,12)& (6,14)& (7,16)& (10,20)& (12,24) \\
\hline
\end{tabular}
\label{tab:ucj_layers}
\end{table}

As mentioned in the main text, we use $T$-count and $T$-depth as the quantities to compare \ac{SQD} and \ac{SQD-AA} with \ac{iQPE}. Thus, we describe how we obtain the $T$-gates for the \ac{UCJ} ansatz in the following. We count the number of non-Clifford single-qubit rotations that are then decomposed to Clifford and $T$-gates using \ac{RUS} synthesis \cite{bocharov_efficient_2015}. A basis rotation $e^K$ comprises $n$ $R_z$ rotations with depth one, where $n$ is the number of qubits, and $(n/2) (n/2 -1)$ Givens rotations with depth $n/2$ \cite{motta_bridging_2023}. The decomposition of a Givens rotation requires two parallel $R_z$ rotations \cite{kivlichan_improved_2020}. The Coulomb operator $e^{\mathrm{i}J}$ consists of $n$ $R_z$ rotations with depth one and $n/2 (n-1)$ controlled phase gates $U_{nn}$ with depth $n$, whose decomposition involves three $R_z$ rotations with depth three \cite{glaser_controlled-controlled-phase_2023}. Thus, the total number of single-qubit rotations $N_{R_z, \mathrm{UCJ}}$ is given by
\begin{flalign}
    N&_{R_z, \mathrm{UCJ}} = L \cdot [ 2 N_{R_z,e^K} + N_{R_z,e^{\mathrm{i}J}}] &\nonumber \\ &= L \cdot \left[ 2 \cdot \left(n + n\left(\frac{n}{2}-1\right) \right) + n +3 \frac{n}{2}\left(n-1\right) \right]&
\end{flalign}
with a corresponding circuit depth $d_{R_z, \mathrm{UCJ}}$ of
\begin{align}
    d_{R_z, \mathrm{UCJ}} &= L \cdot [ 2 d_{R_z,e^K} + d_{R_z,e^{\mathrm{i}J}}] \nonumber \\ &= L \cdot \left[ 2 \cdot \left(1 + \frac{n}{2} \right) + 1 + 3n \right].
\end{align}
Using \ac{RUS} synthesis, the number of $T$-gates required to synthesize a single-qubit rotation $R_z$ within error $\epsilon$ is on average \cite{bocharov_efficient_2015}
\begin{align}
    T_{\mathrm{synth}} = 1.15 \log_2 \left( 1/\epsilon\right) + 9.2.
    \label{eq_app:sk_individual}
\end{align}
Assuming that errors add at most linearly \cite{kivlichan_improved_2020}, we use 
\begin{align}
     N_{T,\mathrm{RUS}} =  1.15 \log_2 \left( N_{R_z, \mathrm{UCJ}}/\epsilon_{\mathrm{tot}} \right) + 9.2
     \label{eq_app:total_sk_ucj}
\end{align}
as the number of $T$-gates required to implement each $R_z$, where we choose $\epsilon_{\mathrm{tot}} = 10^{-4}$. Note that small deviations compared to the full \ac{UCJ} ansatz are tolerable as long as all important bitstrings are still measured with a reasonable number of shots. Moreover, the same synthesized $\Uprep$ and its exact inverse are used in every \ac{AA} step, so the synthesis error does not accumulate. Since $T$-gates in the \ac{RUS} decomposition occur sequentially, $N_{T,\mathrm{RUS}}$ can be multiplied with $N_{R_z, \mathrm{UCJ}}$ and $d_{R_z, \mathrm{UCJ}}$ to obtain $T$-count and $T$-depth of the \ac{UCJ} ansatz, respectively.\\

\begin{figure*}[t]
    \centering
    \includegraphics[width=\textwidth]{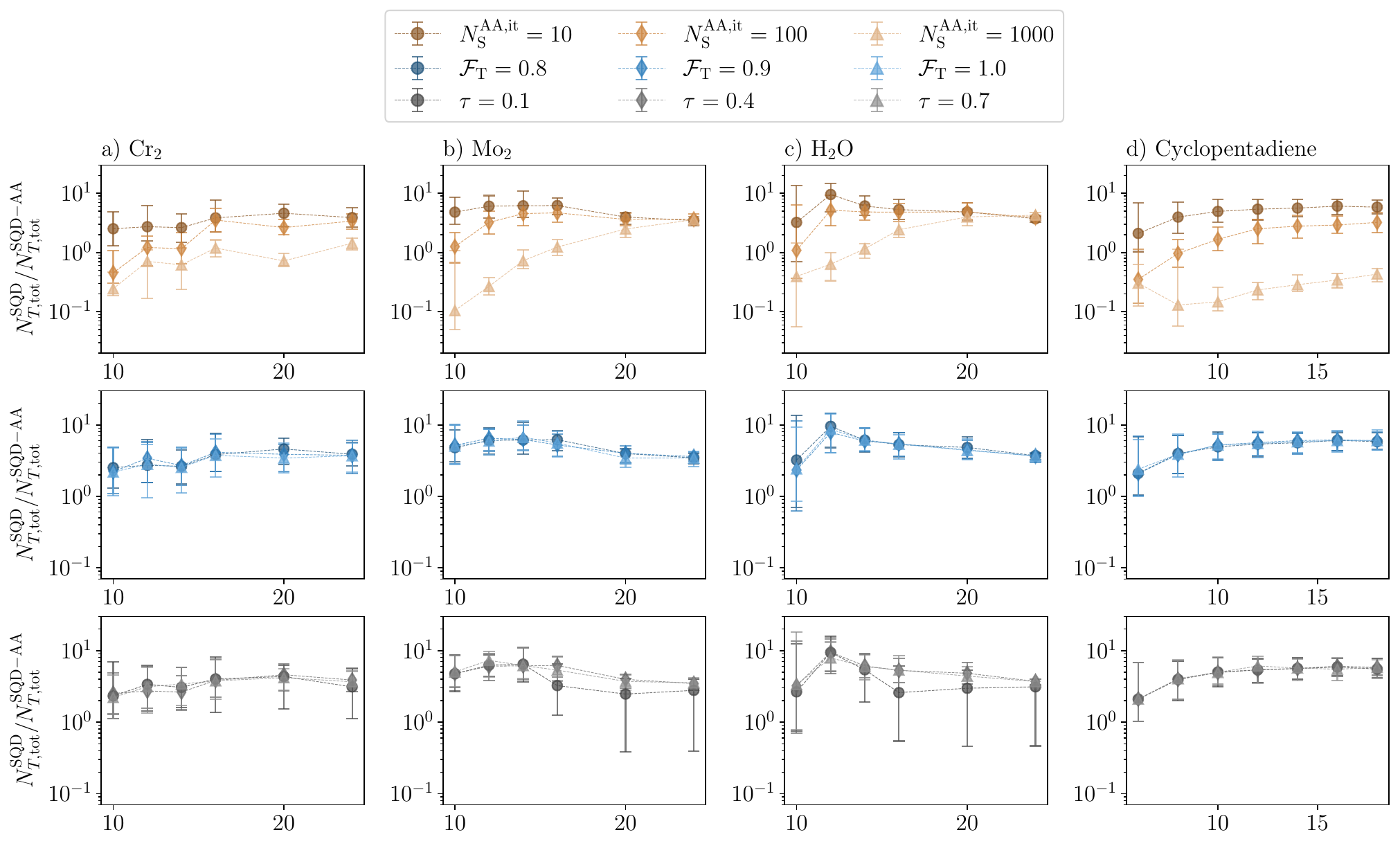}
    \caption{\textbf{Reduction in $\boldsymbol{T}$-gates for different numbers of shots $\boldsymbol{\Ns}$, target fidelities $\boldsymbol{\Ft}$ and threshold parameters $\boldsymbol{\tau}$.} In the first row, the median reduction in $T$-gates (i.e., $N_{T,\mathrm{SQD}} / N_{T,\mathrm{SQD-AA}}$) for different numbers of shots with fixed target fidelity $\Ft=0.8$ and fixed $\tau = 0.4$ is shown for a) Cr$_2$, b) Mo$_2$, c) H$_2$O, and d) cyclopentadiene. In the second row, the median reduction in $T$-gates for fixed $\Ns=10$ and $\tau=0.4$ is presented for different target fidelities and the same systems. In the third row, we vary the threshold $\tau$ (which determines when the algorithm converges) and fix $\Ft=0.8$ and $\Ns=10$. Error bars represent the 68\,\% range of 100 restarts.}
    \label{fig:params}
\end{figure*}

\textit{b) Adiabatic state preparation:}
Under well-known assumptions, one can obtain an approximate ground state of the problem Hamiltonian \(H\) by adiabatically transforming the ground state $\ket{\Psi_Z}$ of a simple initial Hamiltonian \(H_Z\) into an approximate ground state of \(H\). Split the Hamiltonian as
\begin{align}
H = H_Z + H_I,
\end{align}
and introduce the time-dependent Hamiltonian
\begin{align}
H(u) = H_Z + w(u)\,H_I,
\end{align}
where
\begin{align}
H_Z = \sum_{n=N_I+1}^{N_P} a_n P_n
\qquad\text{and}\qquad
H_I = \sum_{n=1}^{N_I} a_n P_n,
\end{align}
with \(H_Z\) containing only single-\(Z\) Pauli terms, \(N_I\) the number of terms in \(H_I\) and $N_P$ the total number of Pauli terms in $H$. Choosing the sweep function so that \(w(0)=0\) and \(w(1)=1\), the protocol adiabatically carries the ground state of \(H(0)=H_Z\) to an approximate ground state of \(H(1)=H_Z+H_I\), provided the two states remain adiabatically connected along the path defined by \(w\).
The time-ordered evolution implementing the adiabatic state preparation is
\begin{align}
\mathcal{A}(T)=\mathcal{T}\exp\!\left(-\imag\int_0^T H\!\left(\frac{t}{T}\right)\,\mathrm{d}t\right),
\end{align}
where \(T\) is the total sweep time, and increasing \(T\) improves the fidelity to the exact ground state of \(H\).

A common approach to realize \(\mathcal{A}(T)\) on a quantum computer is to approximate the continuous evolution with a Trotter-Suzuki decomposition: discretize the time interval into $k$ finite steps,
\begin{align}
    \mathcal{T}\, \mathrm{exp}\lp -\imag \int_0^T H\lp t\rp \mathrm{d}t\rp \approx \prod_{a=1}^{k} \mathrm{exp}\lp -\imag H\lp a \frac{T}{k}\rp \frac{T}{k}\rp,\label{eq_app:time_discretized}
\end{align}
and decompose each exponential of the resulting piecewise-constant Hamiltonians into implementable gate sequences \cite{sun_trotterized_2020,suzuki_decomposition_1985},
\begin{align}
    \exp\left( -\imag u \sum_{h\in H(u)} h \right) &= \nonumber \\\left[\prod_{h\in H(u)} \exp\left( -\imag \frac{u}{m} h \right)\right]^m + \mathcal{O}\lp \frac{u^2}{m} \rp.\label{eq_app:op_discretized}
\end{align}
Here, the number of Trotter repetitions $m$ can be increased to gain algorithmic accuracy at the cost of deeper gate sequences.
While randomized methods circumventing discretization errors exist and have been implemented on quantum hardware, those recently developed methods only implement $\mathcal{A}(T)$ exactly on average\ \cite{granet2024hamiltonian,granet2025practicality,kiumi2025te,nutzel2025ground}.
For this reason the randomized states do not individually resemble the ground state of $H$, and are therefore not suited for sampling in the computational basis as done in this work.
Instead, we will choose a Trotterization of $\mathcal{A}(T)$ as described in \Eq \ \eqref{eq_app:op_discretized}.

More specifically, we have to choose the number of time steps $k$, the number of Trotter repetitions $m$, and the total sweep time $T$.
It is well known that $T$ generally scales with the inverse square of the minimal energy gap $\Delta$, $T\propto \Delta^{-2}$.
For the specific cyclopentadiene system considered here, increasing the system size only adds orbitals to the bath (and not to the system), and therefore $\Delta$ stays roughly constant. 
We thus set $T=2$\,$\mathrm{Ha}^{-1}$ throughout the system sizes considered in this work.

Next, $m$ and $k$ need to be chosen.
To do so, we explore several combinations of $(m,k)$, evaluate the exact resulting statevectors, and calculate their expectation values with respect to $H$.
If an expectation value is closer to the exact ground state energy than the energy expectation value of the initial state $\ket{\Psi_Z}$ by at least 1\,mHa, we add the corresponding pair $(m,k)$ to the set of feasible pairs $\mathcal{P}$.
The best pair $(m^*,k^*)$ is then chosen as
\begin{align}
    (m^*,k^*) = \underset{(m,k)\in \mathcal{P}}{\mathrm{argmin}}\, m\cdot k,
\end{align}
which minimizes the gate count.
This procedure is repeated for each system size.

To obtain the $T$-complexity, we need to consider the number of arbitrary single-qubit $R_z$ rotations that appear when implementing the exponentials in \Eq \ \eqref{eq_app:op_discretized} with the circuit shown in \Fig \ \ref{fig:trott_implement} (note that the control is omitted for the \ac{ASP}). The total number of $R_z$ rotations is given by 
\begin{align}
    N_{R_z, \mathrm{tot}} = N_{P} \cdot m^* \cdot k^*.
\end{align}
These single-qubit rotations are then decomposed with \ac{RUS} synthesis (see \Eq \ \eqref{eq_app:total_sk_ucj}), where $N_{T,\mathrm{RUS}}$ $T$-gates are required for a total circuit synthesis error of $\epsilon_{\mathrm{tot}} =10^{-4}$. Therefore, the $T$-count is $N_{T,\mathrm{RUS}} \times N_{R_z, \mathrm{tot}}$. To obtain the $T$-depth, Pauli gadgets that act on disjoint qubits are executed in parallel.

\subsection{Optimization of Parameters for SQD-AA}\label{subsec:params_sqdaa}
In this section, we determine the ideal number of shots $\Ns$, target fidelity $\Ft$ and threshold $\tau$ for \ac{SQD-AA}. For that, we plot the reduction in the $T$-count compared to \ac{SQD} for different parameters and molecules in \Fig \ \ref{fig:params}.

In the first row of \Fig \ \ref{fig:params}, we vary the number of shots per iteration $\Ns$ at fixed $\Ft=0.8$ and $\tau=0.4$. For all molecules, we find that the reduction in the $T$-count is the highest for $\Ns = 10$. However, with increasing system size, the gap between different numbers of shots tends to decrease and, for example, for H$_2$O with 24 qubits we observe a higher reduction in the $T$-count for $\Ns=100$ and $\Ns=1000$. That is, for small systems often a small number of shots of $\sim 10^3$ is sufficient to sample all important configurations. Therefore, the overhead $\Ns$ to determine the number of steps accurately enough reduces the advantage of \ac{SQD-AA}. Yet, for larger systems, this overhead is smaller in relation to the total number of shots, and at some point a larger $\Ns$ might be beneficial, since the number of steps can then be determined more accurately in this case. Within this work, we stick to $\Ns=10$ for $n\leq20$ and use $\Ns=100$ for $n>20$.

The second row of \Fig \ \ref{fig:params} shows the reduction in $T$-count for different target fidelities $\Ft$ at fixed $\Ns=10$ and $\tau=0.4$. Here, differences are less pronounced. A value of $\Ft=1.0$ often yields the lowest reduction in the $T$-count, while we typically observe the highest reduction for $\Ft=0.8$. Therefore, we choose a target fidelity of $\Ft=0.8$ for simulations of \ac{SQD-AA} throughout this paper.

Finally, different convergence thresholds $\tau$ are evaluated in the third row of \Fig \ \ref{fig:params}, where $\Ft=0.8$ and $\Ns=10$. Here, the reduction in $T$-gates is relatively similar for $\tau=0.4$ and $\tau=0.7$, whereas for $\tau=0.1$ we observe a lower reduction, especially for larger systems. Therefore, we choose the intermediate value of $\tau=0.4$. Note that for an exponentially decaying distribution with $\alpha=1$, $\Delta_{k-1,k}\approx 0.924$ (cf. \Eq \ \eqref{eq:conv_delta}) for any $k$, which is the regime where \ac{SQD-AA} is most efficient.

\subsection{Results for different Molecules}\label{subsec:dif_sys}
\begin{figure}[h]
    \centering
    \includegraphics[width=\columnwidth]{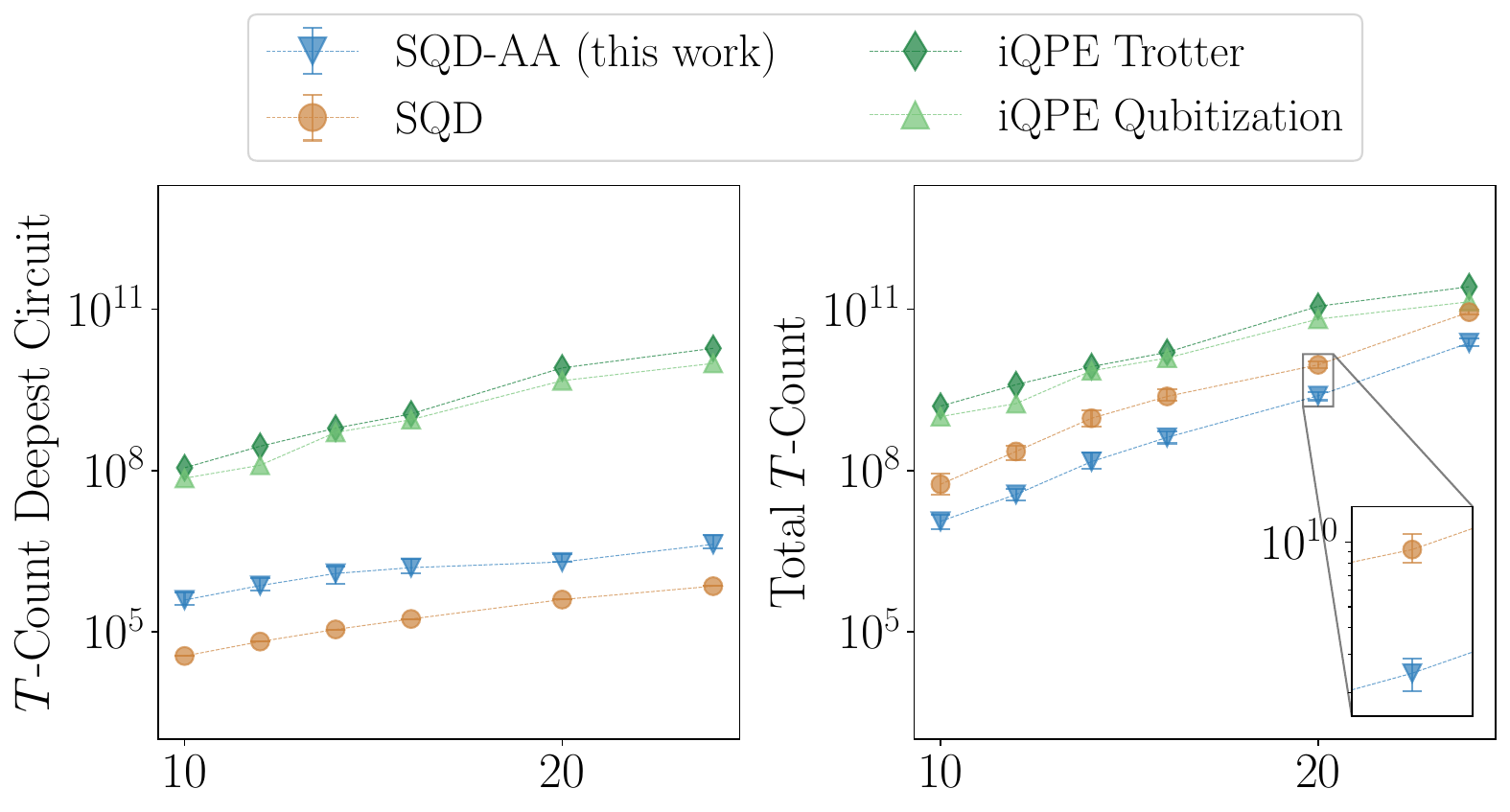}
    \includegraphics[width=\columnwidth]{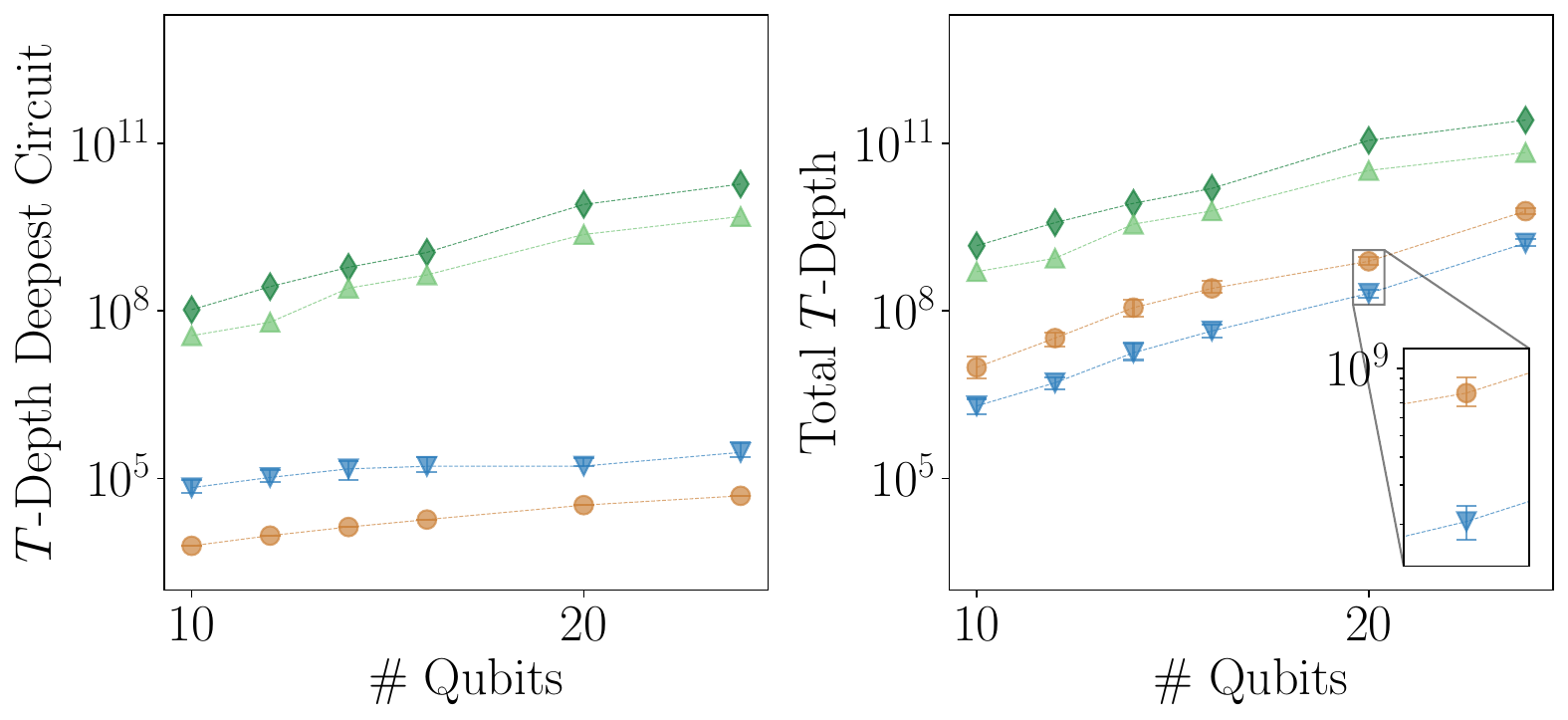}
    \caption{\textbf{$\boldsymbol{T}$-count (upper panels) and $\boldsymbol{T}$-depth (lower panels) to obtain the GSE of Mo$_{\boldsymbol{2}}$ within $\boldsymbol{\epsilon = 1.6 \times 10^{-3}}$\,Ha.} The left panels show $T$-count (upper left panel) and $T$-depth (lower left panel) of the deepest circuit, i.e., the highest number of $T$-gates that are executed within one shot. The right panels display the total $T$-count (upper right panel) and total $T$-depth (lower right panel) which is the $T$-count / $T$-depth multiplied with the total number of shots. We plot median values of 100 repetitions, with error bars representing the 68\,\% range of the restarts. Here, $\Ns=10$ for \mbox{$n\leq20$} and $\Ns=100$ for $n>20$, $\Ft=0.8$, and $\tau=0.4$. The inset shows a zoomed-in view of the $T$-complexity for \ac{SQD-AA} and \ac{SQD} for 20 qubits.}
    \label{fig:mo2}
\end{figure}

In the following, we compare $T$-depth and $T$-count for \ac{SQD}, \ac{SQD-AA} and \ac{iQPE} with Trotterization and qubitization for different molecules. In the main text, the results for Cr$_2$ are shown (see \Fig \ \ref{fig:cr2}). Here, we discuss the same plots for Mo$_2$ and H$_2$O. For Mo$_2$ results are shown in \Fig \ \ref{fig:mo2}.

Overall, we observe similar trends as for Cr$_2$. That is, $T$-depth and $T$-count of the deepest circuit (left panels of \Fig~\ref{fig:mo2}) are several orders of magnitude higher for \ac{iQPE} compared to \ac{SQD} and \ac{SQD-AA}, where the gap increases with system size. Additionally, we obtain the lowest total $T$-count and $T$-depth (right panels of \Fig~\ref{fig:mo2}) for \ac{SQD-AA} with a reduction in the $T$-count of up to a factor of $\sim$ 6 compared to \ac{SQD}. In contrast to Cr$_2$, however, the gap between \ac{SQD}-methods and \ac{iQPE} is relatively small for 24 qubits (upper right panel of \Fig \ \ref{fig:mo2}). Following the trend of the curves, we expect a lower total $T$-count for \ac{iQPE} than \ac{SQD}-methods for larger systems.\\

\begin{figure}[h]
    \centering
    \includegraphics[width=\columnwidth]{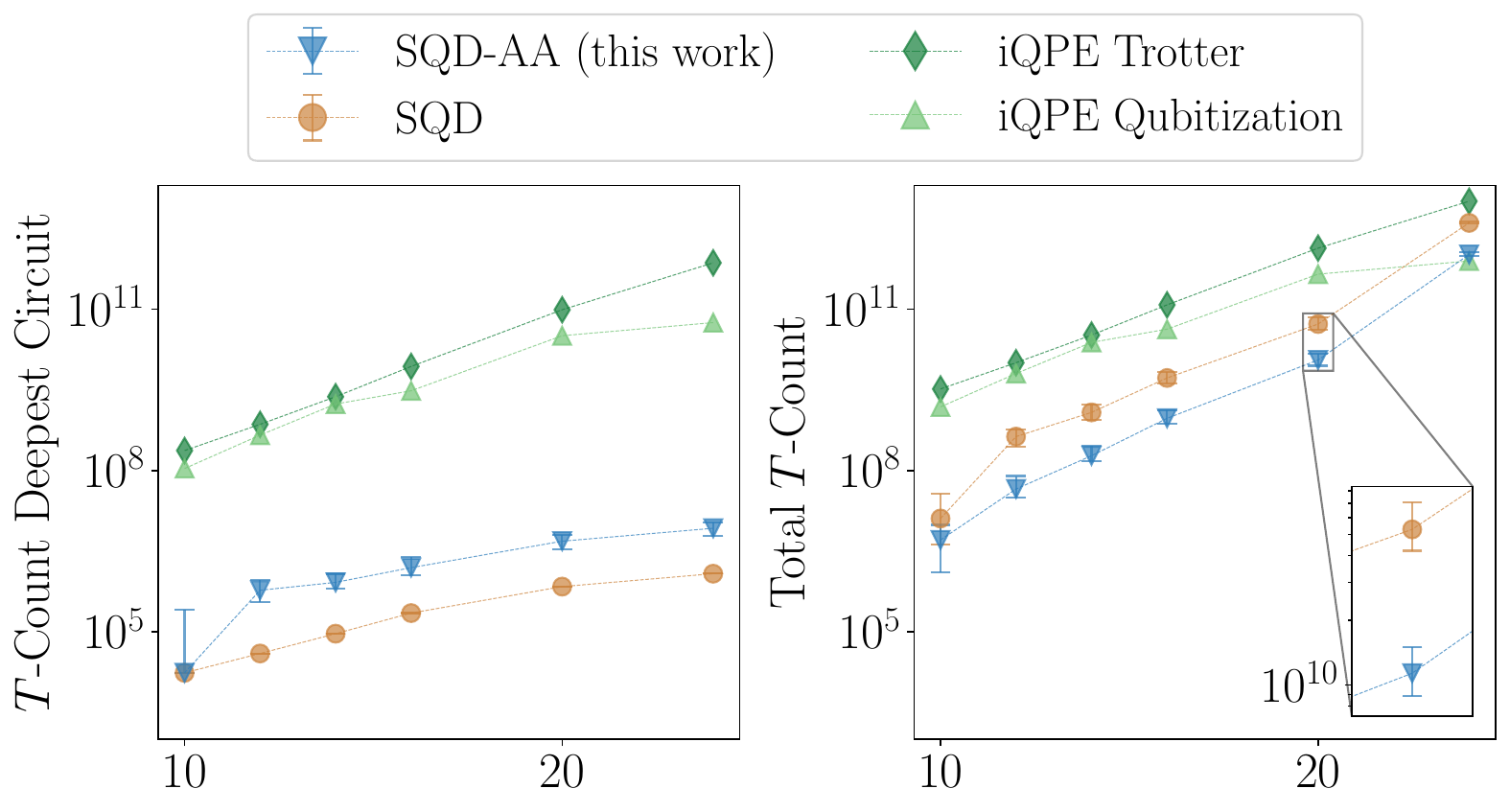}
    \includegraphics[width=\columnwidth]{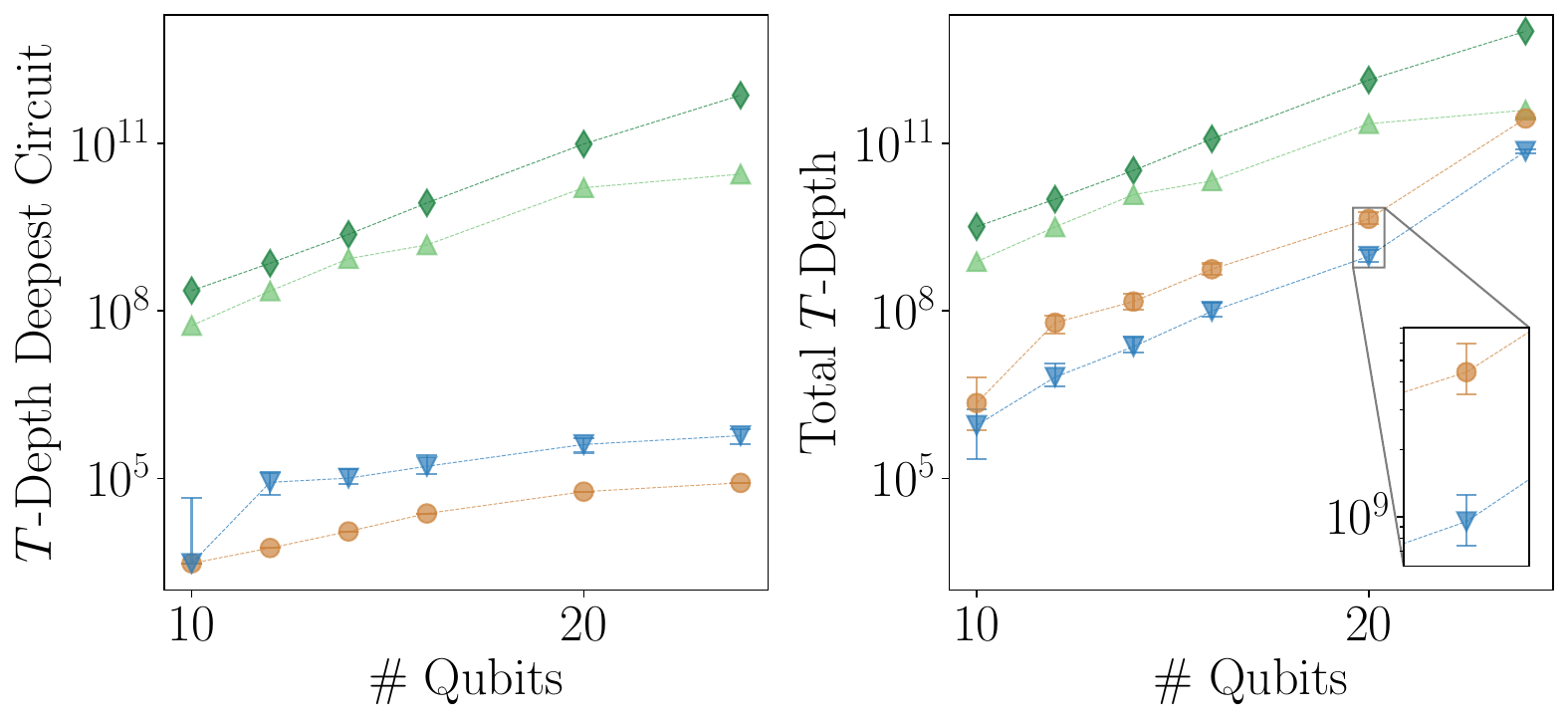} 
    \caption{\textbf{$\boldsymbol{T}$-count (upper panels) and $\boldsymbol{T}$-depth (lower panels) to obtain the GSE of H$_{\boldsymbol{2}}$O within $\boldsymbol{\epsilon = 1.6 \times 10^{-3}}$\,Ha.} The left panels show $T$-count (upper left panel) and $T$-depth (lower left panel) of the deepest circuit, i.e., the highest number of $T$-gates that are executed within one shot. The right panels display the total $T$-count (upper right panel) and total $T$-depth (lower right panel) which is the $T$-count / $T$-depth multiplied with the total number of shots. We plot median values of 100 repetitions, with error bars representing the 68\,\% range of the restarts. Here, $\Ns=10$ for \mbox{$n\leq20$} and $\Ns=100$ for $n>20$, $\Ft=0.8$, and $\tau=0.4$. The inset shows a zoomed-in view of the $T$-complexity for \ac{SQD-AA} and \ac{SQD} for 20 qubits.}
    \label{fig:h2o}
\end{figure}

For H$_2$O (\Fig \ \ref{fig:h2o}) \ac{iQPE} with qubitization outperforms \ac{SQD}-methods already for 24 qubits in the total $T$-count (upper right panel of \Fig \ \ref{fig:h2o}). Additionally, the $T$-depth of \ac{iQPE} with qubitization is already similar to that of \ac{SQD} for this system size (lower right panel of \Fig \ \ref{fig:h2o}). Still, as for Mo$_2$ and Cr$_2$, $T$-count and $T$-depth of the deepest circuit are several orders of magnitude higher for \ac{iQPE} (left panels of \Fig \ \ref{fig:h2o}). Therefore, we argue that in early fault-tolerance, when only a limited number of logical $T$-gates can be executed with sufficiently low logical errors, there is an area where \ac{SQD-AA} can be executed while circuits for \ac{iQPE} would be too deep. Furthermore, note that \ac{iQPE} with qubitization requires significantly more ancillas than system qubits for these system sizes. 

\begin{figure}[h]
    \centering
    \includegraphics[width=\columnwidth]{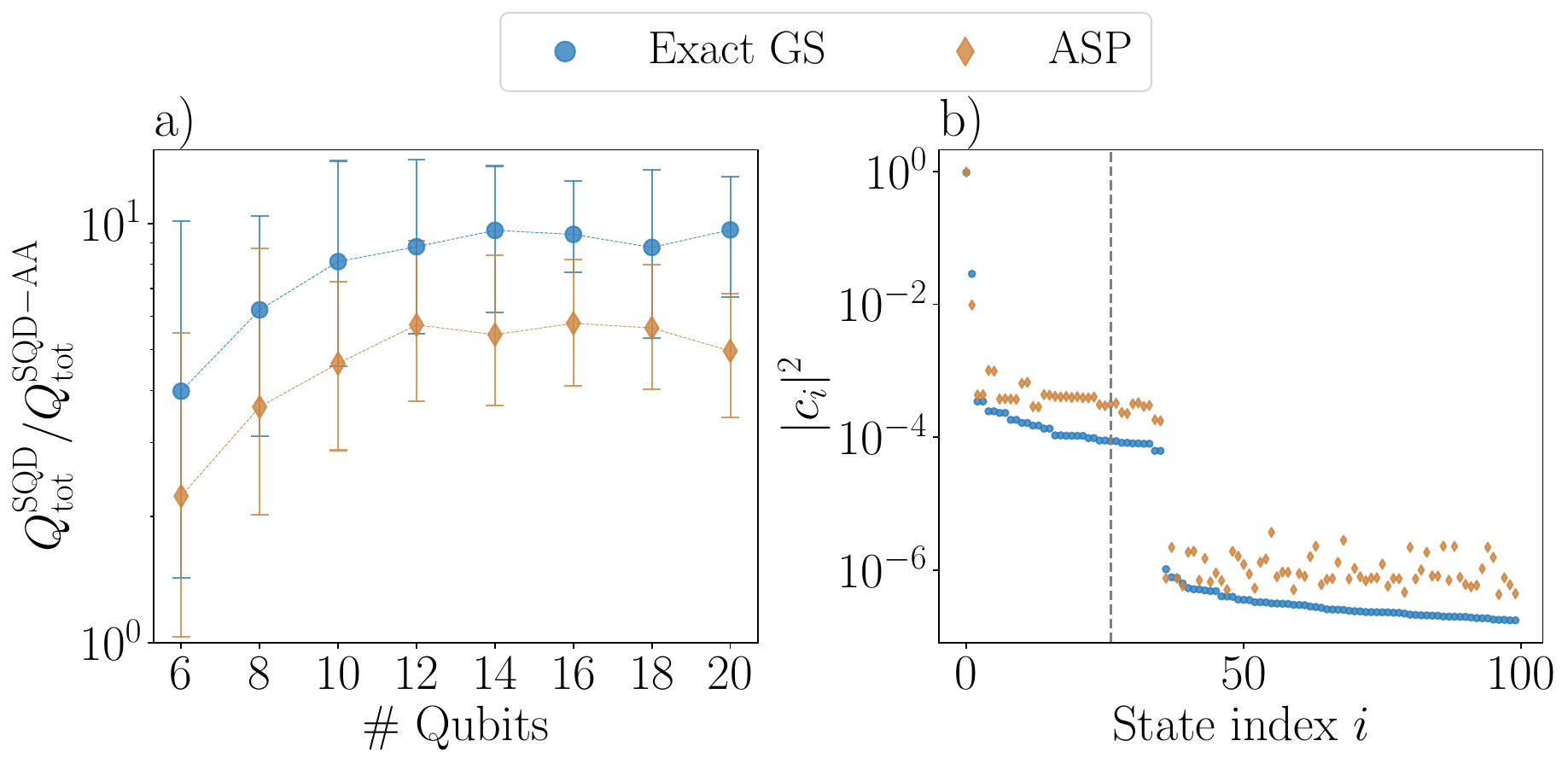} 
    \caption{\textbf{Comparison of the reduction in $\boldsymbol{\Qtot}$ when sampling from the ASP state and the exact ground state of cyclopentadiene. a)} Median reduction in $\Qtot$ for an energy error of $\epsilon = 1.6 \times 10^{-3}$\,Ha. Error bars represent the 68\,\% range of 100 restarts. Here, $\Ns=10$, $\Ft=0.8$, and $\tau=0.4$. \textbf{b)} Probabilities $|c_i|^2$ of computational basis states ordered by decreasing magnitude for 20 qubits. The gray dashed line indicates the number of bitstrings required for an error $\epsilon$ in the GSE. Note that we do not plot the full state, as less probable configurations are not important here.}
    \label{fig:gain_comp_hqs}
\end{figure}

Compared to \ac{SQD}, we observe a reduction in the $T$-count of up to a factor of $\sim$ 10 (right panels of \Fig \ \ref{fig:h2o}). We want to emphasize that (according to Section \ref{subsec:dif_model_dists}) we expect that a greater reduction in the $T$-count is in principle possible. As an example, we compare the reduction in $\Qtot$ (i.e., in the total number of times the state preparation unitary is applied) when sampling from the exact ground state and the adiabatically prepared state of cyclopentadiene in \Fig \ \ref{fig:gain_comp_hqs}.

In \Fig \ \ref{fig:gain_comp_hqs} a) we observe that for both states, the reduction in $\Qtot$ is roughly increasing with system size; however, when sampling from the exact ground state, the reduction in $\Qtot$ is higher. Here, we achieve an improvement of up to a factor of $\sim$ 10, compared to a maximal reduction of a factor of $\sim$ 6 for the adiabatically prepared state. The lower reduction for the adiabatically prepared state is caused by the fact that probabilities of required bitstrings are larger in this case, as can be seen in \Fig \ \ref{fig:gain_comp_hqs} b). Therefore, fewer shots are required for direct sampling and the overhead $\Ns$ for \ac{SQD-AA} is weighted more heavily. Still, this is related to the specific problem and state preparation method. That is, with another state preparation method, the runtime reduction when using \ac{SQD-AA} might be higher. Of course, we cannot make predictions for system sizes where quantum advantage could be achieved, yet, we expect that \ac{SQD-AA} performs better, when a higher number of shots is required to measure all necessary bitstrings, which is usually the case for larger systems.

\section{Iterative Quantum Phase Estimation (iQPE) \label{sec:appC}}
In this work, we use \ac{iQPE} as benchmark to determine the \ac{GSE}, as it is considered among the most efficient quantum algorithms for this task \cite{martyn_grand_2021}. We use the iterative version of \ac{QPE}, since it requires shorter circuits than standard \ac{QPE} and is therefore more suitable for early fault-tolerant devices \cite{dobsicek_arbitrary_2007, smith2022iterative}. There exist several modifications of \ac{iQPE} that can lower the resource requirements. Using adaptive phase estimation techniques, the total cost can be reduced by a factor of 2.63 \cite{kivlichan_improved_2020}. Further improvements are possible with Bayesian phase estimation, where a factor of 3.82 can be achieved \cite{wiebe_efficient_2016}. The ultimate lower bound would correspond to an improvement by a factor of 4, however, at the cost of multiple control qubits \cite{babbush_encoding_2018}. Within this work, we argue that these algorithms could not close the gap of several orders of magnitude between \ac{SQD}-methods and \ac{iQPE} in the $T$-complexity of the deepest circuit, as can be seen for example in Figs.~\ref{fig:hqs_Tgates} and~\ref{fig:cr2}. Thus, we briefly review \ac{iQPE} \cite{dobsicek_arbitrary_2007} and then describe explicit methods to implement the Hamiltonian $H$ as a unitary, that is, via Trotterization and qubitization. Additionally, we explain how we obtain $T$-count and $T$-depth for both algorithms.

The aim of \ac{iQPE} is to compute eigenvalues $e^{2\pi i \phi}$ of a unitary $U$ up to desired precision $\epsilon$ on a quantum computer. 
Thus, to obtain the \ac{GSE} of a Hermitian operator $H$, it must first be embedded into a unitary so that the phases can be mapped to eigenvalues of $H$, $E = f(\phi)$. 
For now, we assume that the Hamiltonian is embedded in a unitary, and describe explicit constructions in subsequent sections. The phase $\phi$ can be estimated bit-wise up to $m$ bits $\phi \approx 0.\phi_1 \phi_2 \ldots \phi_m$, where $0.\phi_1 \phi_2 \ldots \phi_m = \sum_{k=1}^m 2^{-k} \phi_k$ is a binary expansion. The circuit for \ac{iQPE} is illustrated in \Fig \ \ref{fig:iqpe_circ}.

\begin{figure}
    \centering
    \begin{quantikz}
    \lstick{$\ket{0}$} & \gate{H} & \ctrl{1} & \gate{R_z(\omega_k)} & \gate{H} & \meter{\phi_k} \\
    \lstick{$\ket{\tilde{\Psi}_{\mathrm{GS}}}$} &\qwbundle{n}& \gate{U^{2^{k-1}}} &&&
    \end{quantikz}
    \caption{\textbf{Circuit that implements the $\boldsymbol{k}$th iteration of \ac{iQPE}.} The angle $\omega_k = -2\pi(0.0\phi_{k+1}\phi_{k+2} \ldots \phi_m)$ depends on previous iterations and $\omega_m=0$.}
    \label{fig:iqpe_circ}
\end{figure}

Here, the upper line represents an ancilla that is measured, while the lower line corresponds to physical qubits. At this point, we assume that the exact ground state $\ket{\Psi_{\mathrm{GS}}}$ of $U$ can be prepared. This can be readily generalized to approximate ground states $\GS$. Furthermore, we first assume that the phase is an exact binary i.e., $\phi = 0.\phi_1 \phi_2 \ldots \phi_m$. Starting with the least significant bit $k=m$, a controlled-$U^{2^{k-1}}$ gate is applied to the physical qubits. This results in the state $\frac{1}{2} [(1+e^{\mathrm{i}2\pi 0.\phi_m}) \ket{0} + (1-e^{\mathrm{i}2\pi 0.\phi_m}) \ket{1}] \ket{\Psi_{\mathrm{GS}}}$ before measurement. The probability to measure $0$ is thus $\cos^2(\pi(0.\phi_m))$ which is one for $\phi_m=0$ and zero else. Hence, $\phi_m$ can be extracted deterministically. In further iterations ($k=m-1, \ldots, 1$) we proceed similarly, but with an additional rotation $R_z(\omega_k)$. That is, in the second iteration the phase after applying $U^{2^{m-2}}$ is $0.\phi_{m-1} \phi_{m}$. To deterministically extract the second bit, however, the phase must be reduced to $0.\phi_{m-1}$. We therefore apply a corrective rotation $R_z(\omega_{m-1})$ with $\omega_{m-1}=-2\pi (0.0\phi_m)$, which removes the contribution of the previously determined bit. Repeating this procedure allows all bits to be extracted deterministically \cite{dobsicek_arbitrary_2007}.

Of course, the exact ground state is usually not known, and the phase is often not an exact binary. As a consequence, repeated measurements are required. If an approximate ground state is prepared, it can be expressed in the eigenbasis, $\GS = c_{\mathrm{GS}} \ket{\Psi_{\mathrm{GS}}} + \sum_k c_k \ket{\lambda_k}$. Hence, the probability to measure the ground state phase is $|c_{\mathrm{GS}}|^2$ \cite{cruz_optimizing_2020}. In case the phase cannot exactly be expressed as a binary expansion, there is a remainder $\phi = 0.\phi_1 \phi_2 \ldots \phi_m + \delta 2^{-m}$ where $0 \leq \delta < 1$. Accordingly, the probability to extract $0.\phi_1 \phi_2 \ldots \phi_m$ is
\begin{align}
    P(\delta) = \prod_{k=1}^m P_k = \frac{\sin^2({\pi \delta})}{2^{2m} \sin^2(\pi 2^{-m} \delta)}
\end{align}
where $P_k = \cos^2(\pi 2^{k-m-1}\delta)$. For an accuracy of $2^{-m}$ we accept rounding up and down and the success probability $P_{\mathrm{bin}}(\delta) = P(\delta) + P(1-\delta)$ is lower bounded by $8/\pi^2$ independent of $m$ \cite{dobsicek_arbitrary_2007}. Therefore, the overall probability to extract the phase is $P_{\mathrm{bin}}(\delta)\cdot |c_{\mathrm{GS}}|^2 \geq 8|c_{\mathrm{GS}}|^2/\pi^2 $. In this work, we estimate the measurement shots such that the ground state phase is determined by a majority vote, by considering each shot as an independent Bernoulli trial. Moreover, we use the same initial state as for \ac{SQD} and \ac{SQD-AA} to enable a fair comparison. Having reviewed the general concepts of \ac{iQPE}, we now proceed to describe the explicit methods to embed the Hamiltonian in a unitary and estimate the $T$-gates of the circuits.

\subsection{iQPE with Trotterization}\label{subsec:iqpe_trotter}
\begin{figure}[htbp]
    \centering
    \begin{minipage}{0.49\textwidth}
        \centering
        \includegraphics[width=0.48\textwidth]{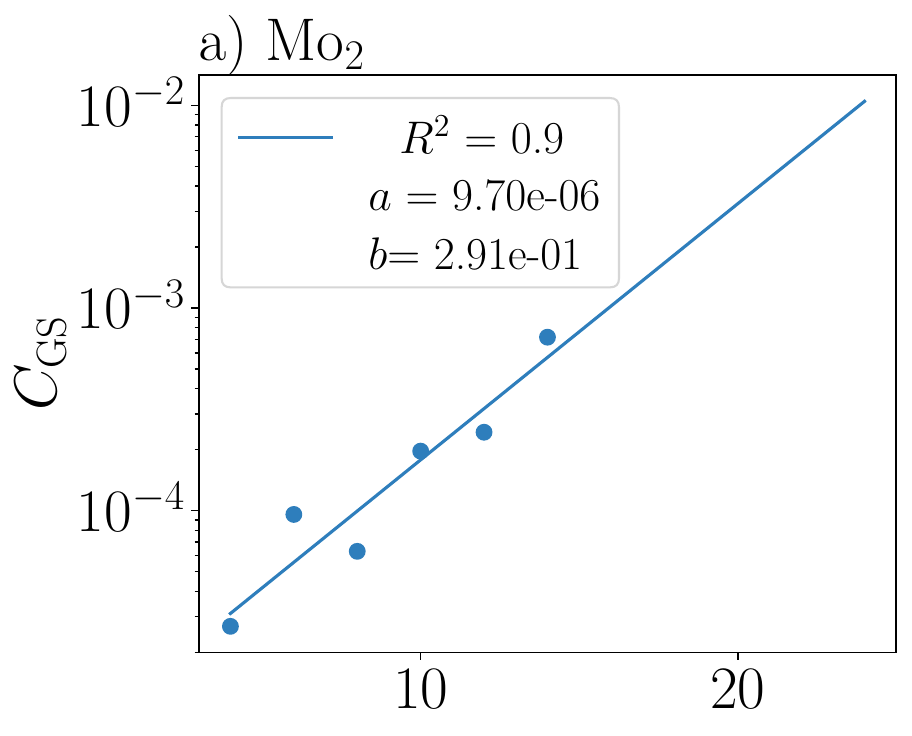} \includegraphics[width=0.48\textwidth]{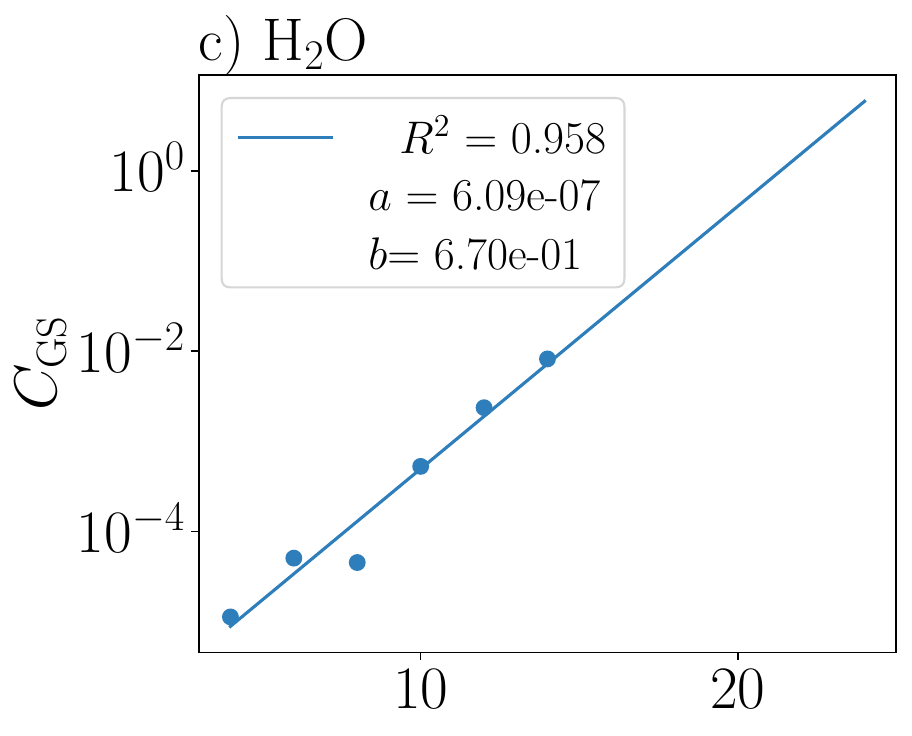}
    \end{minipage}
    
    \vspace{0.2cm}

    \begin{minipage}{0.49\textwidth}
        \centering
        \includegraphics[width=0.48\textwidth]{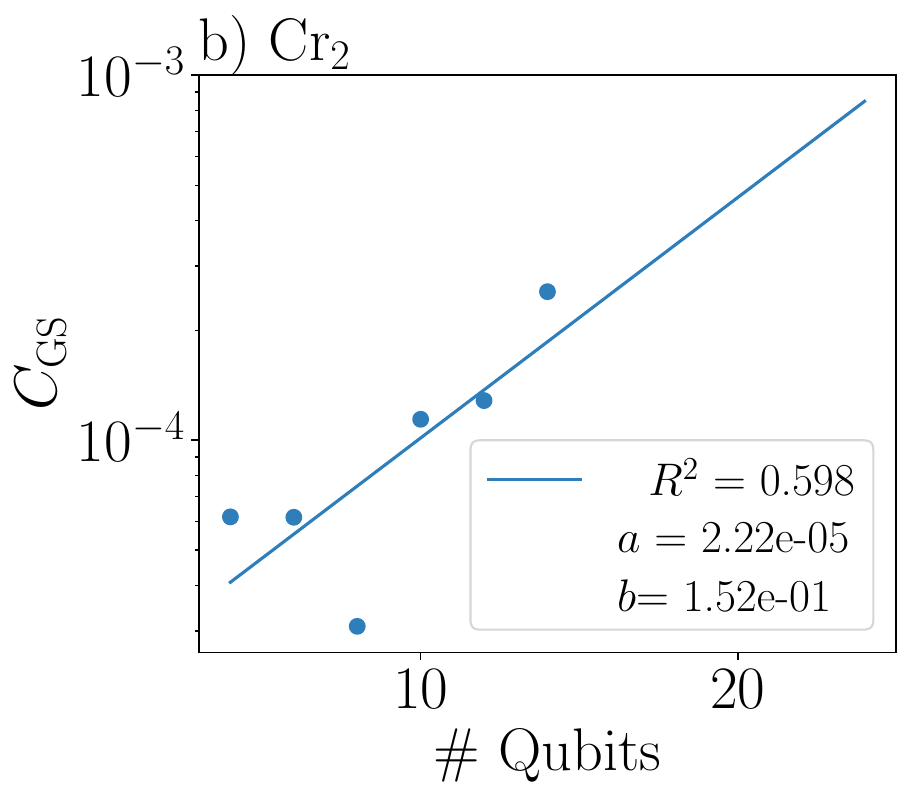} \includegraphics[width=0.48\textwidth]{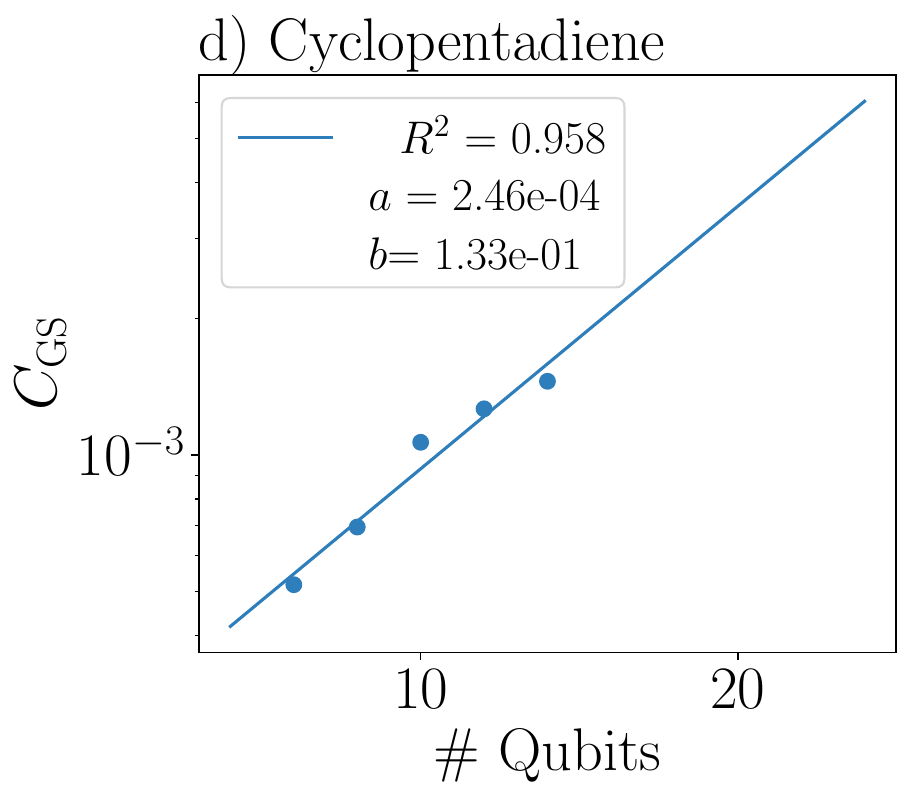}
    \end{minipage}
    \caption{\textbf{Numerically determined Trotter error constants $\boldsymbol{C_{\mathrm{GS}}}$.} The exact energy errors $\Delta E_{\mathrm{TS}}$ are determined for different time steps $\Delta\tau \in [10^{-4},10^{-1}]$ and $C_{\mathrm{GS}}$ is the average of the respective constants (which are in all cases nearly identical). For larger systems, we extrapolate the error constants by fitting an exponential of the form $f(x) = a \exp(bx)$. Results are shown for a) Mo$_2$, b) Cr$_2$, c) H$_2$O and d) cyclopentadiene.}
    \label{fig:error_constants}
\end{figure}

The natural choice to implement $H$ as a  unitary is via an exponential of the form
\begin{align}
    U = e^{-\mathrm{i}Ht}.
\end{align}
In this case, $\exGSE = -2\pi \phi_{\mathrm{GS}}/t$. In the \ac{JW} representation, the Hamiltonian is expressed as a sum of Pauli strings $P_i$, $H = \sum_{i=1}^{L} c_i P_i$, where we use a lexicographic ordering. Since the individual terms generally do not commute, a common approach to implement $U$ on a quantum computer is the second-order Trotter formula
\begin{align}
e^{-\mathrm{i} H t} \approx
\left[ \left(\prod_{i=1}^{L} e^{-\mathrm{i} P_i c_i \, t/2s}\right) \left(\prod_{i=L}^{1} e^{-\mathrm{i} P_i c_i\, t/2s}\right) \right]^s \nonumber \\
\equiv e^{-\mathrm{i} H_{\mathrm{eff}} t}
\end{align}
where $s$ is the number of Trotter steps \cite{poulin_trotter_2014}. The difference in the \ac{GSE} of the effective Hamiltonian $H_{\mathrm{eff}}$ and the exact \ac{GSE} is bounded by
\begin{align}
    \Delta E_{\mathrm{TS}} = |\exGSE - E_{\mathrm{GS,eff}}| \leq C_{\mathrm{GS}} \Delta \tau^2,
\end{align}
with $\Delta \tau = t/s$. The error constant $C_{\mathrm{GS}}$ can, for example, be derived using nested commutator norm bounds. However, for larger systems this approach is computationally expensive and the resulting bounds are relatively loose \cite{childs_theory_2021}. Thus, we follow a numerical approach introduced in Appendix D of Günther et al. \cite{gunther_phase_2025}. That is, we determine the energy error $\Delta E_{\mathrm{TS}}$ for different time steps $\Delta \tau$ and fit a power law to obtain $C_{\mathrm{GS}}$. With increasing system size, however, it becomes computationally expensive to calculate $H_{\mathrm{eff}}$ through the logarithm of the exponential. Hence, we extrapolate the error constants for systems with more than 14 qubits where we fit the exponential $f(x) = a \exp(bx)$, with fit parameters $a$ and $b$. The error constants and corresponding fits for the molecules used in this work are plotted in \Fig \ \ref{fig:error_constants}.

For all systems but Cr$_2$ we achieve residuals $R^2 \geq 0.9$. For Cr$_2$, we observe a large deviation for eight qubits, where the error constant is even smaller than for four and six qubits. Yet, overall the fits seem reasonable, as we only use the error constants to estimate the $T$-counts. Note that the numerical error constants do not constitute rigorous bounds; however, we still expect them to be sufficient to determine the order of magnitude of the $T$-complexity. Moreover, $T$-counts estimated with constants derived by commutator norm bounds are often larger and overestimated \cite{childs_theory_2021}.

When implementing \ac{iQPE} with Trotterization, the total energy error arises from multiple sources. That is, the Trotterization error, $\Delta E_{\mathrm{TS}}$, the finite precision error of \ac{iQPE}, $\Delta E_{\mathrm{iQPE}}$, and an additional gate-synthesis error $\Delta E_{\mathrm{RUS}}$ introduced when decomposing arbitrary single-qubit rotations in terms of Clifford and $T$-gates with \ac{RUS} synthesis. These errors are not independent and can interact in a nontrivial way. Hence, to obtain the minimal $T$-count (and $T$-depth), we follow the approach of Kivlichan et al. \cite{kivlichan_improved_2020} and minimize the number of $T$-gates under the constraint that
\begin{align}
    0 <  \Delta E_{\mathrm{TS}} + \Delta E_{\mathrm{iQPE}} + \Delta E_{\mathrm{RUS}} \leq 1.6\,\mathrm{mHa}
\end{align}
since the errors add at worst linearly. The total number of $T$-gates is then given by the $T$-count of the Trotterization (i.e., number of non-Clifford single-qubit rotations $N_{\mathrm{rot}, \mathrm{TS}}$ times the \ac{RUS} overhead $N_{\mathrm{RUS}}$) multiplied with the number of times the Trotter unitary is applied in \ac{iQPE} ($N_{\mathrm{rep},\mathrm{iQPE}}$),
\begin{align}
    N_T = N_{\mathrm{rot}, \mathrm{TS}} \cdot N_{\mathrm{RUS}} \cdot N_{\mathrm{rep},\mathrm{iQPE}}.
\end{align}
We now examine the individual contributions. The Trotter approximation consists of products of exponentials of Pauli strings, $e^{-\mathrm{i}P_i c_it/2s}$. Within \ac{iQPE}, these exponentials are controlled by an ancilla qubit. Each such controlled exponential can be realized using Pauli gadgets, shown in \Fig \ \ref{fig:trott_implement} for the example $e^{-\mathrm{i}c_it/2s X_1Y_2Z_3}$ \cite{mansky_decomposition_2023}.

\begin{figure}
    \centering
    \scalebox{0.85}{
    \begin{quantikz}[column sep=0.3cm]
        &&&&& \ctrl{3} &&&&&\\
        &\gate{H} && \ctrl{1}&& && \ctrl{1} && \gate{H} & \\
        &\gate{S^{\dagger}} & \gate{H}& \targ{} & \ctrl{1}&& \ctrl{1} & \targ{} & \gate{H} &\gate{S} &\\
        &&&& \targ{} & \gate{R_z(c_i t/s)} & \targ{} &&&&
    \end{quantikz}}
    \caption{\textbf{Circuit that implements the controlled exponential $\boldsymbol{e^{-\mathrm{i}c_it/2s X_1Y_2Z_3}}$.} The first qubit represents the ancilla of \ac{iQPE} that controls the unitary \cite{mansky_decomposition_2023}.}
    \label{fig:trott_implement}
\end{figure}
 
Since all gates but the $R_z$ rotations are Clifford, we only need to consider those when evaluating the $T$-count. For second-order Trotterization with $s$ steps the total number of $R_z$ rotations is
\begin{align}
    N_{\mathrm{rot}, \mathrm{TS}} = \underbrace{2}_{\mathrm{2. \ order}} \cdot L \cdot s \cdot \underbrace{2}_{\mathrm{controlled}}.
\end{align}
The last factor of two accounts for the fact that two $R_z$ rotations are required to implement one controlled $R_z$ rotation \cite{haner_software_2018}. Moreover, $L$ is the number of Pauli strings in $H$. The number of $T$-gates to synthesize $N_{\mathrm{rot}, \mathrm{TS}}$ $R_z$ rotations with a desired error $\Delta E_{\mathrm{RUS}}$ in the \ac{GSE} estimate is given by 
\begin{align}
    N_{\mathrm{RUS}} \approx 1.15 \log_2\left(\frac{4L \cdot 2 \pi }{\Delta E_{\mathrm{RUS}} \Delta \tau}\right) + 9.2,
\end{align}
where we again assume that errors in individual rotations (see \Eq \ \eqref{eq_app:sk_individual}) add at most linearly \cite{kivlichan_improved_2020}.

Having described how we obtain the $T$-count for Trotterization, we finally evaluate the number of times these exponentials are applied in \ac{iQPE}. The total number of times the controlled unitary is applied is given by
\begin{align}
    \sum_{k=0}^{m-1} 2^k = 2^{m} -1 \approx 2^m \equiv N_{\mathrm{rep, iQPE}}.
\end{align}
The energy error resulting from the finite precision of the phase-estimation bits can be described as
\begin{align}
    \Delta E_{\mathrm{iQPE}} = \frac{2\pi}{t} \Delta \phi \approx \frac{2\pi}{s \Delta \tau} 2^{-m}.
\end{align}
Therefore, by expressing $N_{\mathrm{rep, iQPE}}$ in terms of \ac{iQPE} and Trotter error, we obtain
\begin{align}
    2^m = \frac{2 \pi}{\Delta E_{\mathrm{iQPE}} \Delta \tau \cdot s} \approx \frac{2 \pi \sqrt{C_{\mathrm{GS}}}}{\Delta E_{\mathrm{iQPE}} \sqrt{\Delta E_{\mathrm{TS}}} \cdot s}.
\end{align}
With that, we define the cost function
\begin{flalign}
    C(s, &\Delta \tau, m, \Delta E_{\mathrm{RUS}}) &\notag\\
    &\approx 4 L \left( 1.15 \log_2\left(\frac{4 L \cdot 2 \pi \sqrt{C_{\mathrm{GS}}}}{\Delta E_{\mathrm{RUS}} \sqrt{\Delta E_{\mathrm{TS}}(\Delta \tau)}}\right) + 9.2 \right) \notag &\\ &\hspace{1cm}\cdot \frac{2 \pi \sqrt{C_{\mathrm{GS}}}}{\Delta E_{\mathrm{iQPE}}(m,\Delta \tau, s) \sqrt{\Delta E_{\mathrm{TS}}(\Delta \tau)}},&
\end{flalign}
i.e., the total number of $T$-gates that we minimize under the constraint that the total energy error is below a certain threshold. The optimizations are performed using the \texttt{COBYLA} algorithm, a derivative-free constrained optimizer. Finally, we add the $T$-count of the initial state multiplied by the number of \ac{iQPE} iterations to the optimized value.To obtain the $T$-depth, controlled Pauli gadgets that act on disjoint system qubits are executed in parallel, since they are coupled to the control qubit only via commuting CNOT gates. Moreover, to obtain the total $T$-complexity, the respective quantities are multiplied with the measurement shots estimated such that the ground state phase is obtained with a majority vote. The results of the optimizations are plotted in the main figures, where we compare $T$-counts for \ac{GSE} estimation with different methods.

\subsection{iQPE with Qubitization}\label{subsec:iqpe_qubitization}
Another approach for encoding eigenvalues of a Hamiltonian $H$ exactly into a unitary $Q$ is qubitization \cite{martyn_grand_2021}. The corresponding qubitization unitary $Q$ is defined as
\begin{align}
    Q = \mathrm{PSP}_H \cdot (2|0\rangle \langle 0 | - \mathbb{I}).
\end{align}
Here, $\mathrm{PSP}_H$ denotes a \ac{LCU} representation of $H$ which we describe below. Since $H$ can be expressed as sum of Pauli strings $H= \sum_{i=1}^L c_i P_i$, where each $P_i$ is a unitary, the Hamiltonian can be implemented on a quantum computer by applying appropriate states and controlled operations. First, we define a prepare (PREP) operator that generates the superposition
\begin{align}
    \prep \ket{0} = \sum_{i=1}^L \sqrt{\frac{|c_i|}{\lambda}} \ket{i}
\end{align}
where $\lambda = \sum_i |c_i|$. This operator prepares states with amplitudes corresponding to the absolute values of the coefficients $|c_i|$. The subsequent select (SEL) operator
\begin{align}
    \sel \equiv \sum_{i=1}^L |i\rangle \langle i| \otimes P_i \\
    \sel \ket{i} \ket{\Psi} = \ket{i} P_i \ket{\Psi}
\end{align}
then applies the Pauli string associated with each ancilla state. For negative coefficients $c_i$, the minus sign can be absorbed into the phase of $P_i$. Afterward, $\prep^{\dagger}$ is applied to uncompute the $\prep$ operation. The number of ancillas required depends on the number of Pauli strings in $H$ and is given by $\lceil \log_2 L \rceil$. Therefore, the state after applying PSP$_H$ is
\begin{align}
    \ket{\Psi} &= \mathrm{PSP}_H \ket{0}^{\otimes \lceil \log_2 L \rceil} \GSex \nonumber \\ &= \mathrm{PREP}^{\dagger} \ \mathrm{SEL} \ \mathrm{PREP} \ket{\boldsymbol{0}} \GSex \nonumber \\ &= \frac{\exGSE}{\lambda} \ket{\boldsymbol{0}} \GSex + \sqrt{1- \left( \frac{\exGSE}{\lambda} \right)^2} \ket{\psi^{\perp}}.
    \label{eq_app:qubiti_subspace}
\end{align}

\begin{figure}
    \centering
    \scalebox{0.95}{
    \begin{quantikz}[column sep=0.35cm]
    \lstick{$\ket{0}$} & \gate[2]{\prep} & \octrl{2} \gategroup[3,steps=4,style={dashed,rounded corners,fill=gray!25, inner xsep=2pt},background,label style={label
			position=above,anchor=north,yshift=+0.3cm}]{{SEL}}  & \octrl{1} & \ctrl{1} & \ctrl{2} & \gate[2]{\prep^\dagger} & \\
    \lstick{$\ket{0}$} && \octrl{1} & \ctrl{1} & \octrl{1} & \ctrl{1} &&\\
    \lstick{$\ket{\Psi}$} && \gate{P_1} & \gate{P_2} & \gate{P_3} & \gate{P_4} &&
    \end{quantikz}}
    \caption{\textbf{Example of a circuit that implements $\mathbf{PSP}_H$.} Here, the circuit corresponds to a Hamiltonian with four terms, $H= \sum_{i=1}^4 c_i P_i$.}
    \label{fig:lcu}
\end{figure}

Moreover, the circuit implementing $\mathrm{PSP}_H$ is illustrated in \Fig \ \ref{fig:lcu} for a small example with $L=4$. Writing the circuit-unitary as a matrix, we can see that $H$ is encoded in a block of $\mathrm{PSP}_H$,
\begin{align}
    \mathrm{PSP}_H =
    \begin{bmatrix}
    H/\lambda  & \cdot \\
    \cdot & \cdot
    \end{bmatrix}.
\end{align}
Additionally, one can check that $\mathrm{PSP}_H^2 = \mathbb{I}$, i.e., the unitary represents a reflection. This reflection itself, however, does not yet encode the eigenvalues of $H$. To achieve this, an additional reflection $2|\boldsymbol{0}\rangle \langle \boldsymbol{0} | - \mathbb{I}$ is applied beforehand. These two reflections then form a rotation that encodes the eigenvalues. To see this, we analyze the action of $Q$ on the state in \Eq \ \eqref{eq_app:qubiti_subspace}. We choose this state, since applying $Q$ to $\ket{\boldsymbol{0}} \GSex$ has the same effect as applying PSP$_H$ once, but for higher powers of $Q$ the action differs since $Q$ is a rotation while PSP$_H$ is a reflection.

As can be seen in \Eq \ \eqref{eq_app:qubiti_subspace}, PSP$_H$ defines a two-dimensional subspace as for \ac{AA} (see Section \ref{sec:aa}, \Fig \ \ref{fig:aa_sub}). Here, $\ket{\Psi}$ forms an angle of $\theta = \arccos(\exGSE/\lambda)$ relative to the axis defined by $\ket{\boldsymbol{0}} \GSex$. The operator $2|\boldsymbol{0}\rangle \langle \boldsymbol{0}|- \mathbb{I}$ thus  introduces a reflection about $\ket{\boldsymbol{0}} \GSex$. Subsequently, PSP$_H$ reflects the previous state about an axis that bisects $\ket{\Psi}$ and $\ket{\boldsymbol{0}} \GSex$. That is, these two reflections produce a rotation of $\ket{\Psi}$ by an angle $\theta$. Since the eigenvalues of the rotation operator $Q$ are $e^{\pm i\theta}$ and $\theta = \arccos(\exGSE/\lambda)$, we can exactly encode the \ac{GSE} of $H$ in $Q$. The initial state $\ket{\boldsymbol{0}} \GSex$ can be expressed as equal superposition state of the eigenstates of $Q$,
\begin{align}
    \ket{\boldsymbol{0}} \GSex = \frac{1}{\sqrt{2}} ( \ket{\lambda^+} + \ket{\lambda^-}),
\end{align}
with $\ket{\lambda^{\pm}} = \frac{1}{\sqrt{2}} (\ket{\boldsymbol{0}} \GSex \mp \mathrm{i} \ket{\psi^{\perp}})$. Therefore, the phases $\pm \theta$ are measured with equal probability; however, they both yield the same energy as the cosine is symmetric.

Next, we discuss how we obtain the $T$-complexity when implementing \ac{iQPE} with qubitization. Here, we follow the approach of Babbush et al. \cite{babbush_encoding_2018}. The controlled SEL circuit can be rewritten using so-called unary iterations that are described in detail in Section III.A of Ref. \cite{babbush_encoding_2018}. Using this technique, a sequence of multi-controlled CNOT gates is implemented by computing and uncomputing AND operations. The main advantage is that the uncomputation of AND operations does not require any $T$-gates \cite{gidney_halving_2018}. For the computation of AND operations, we assume a $T$-count of 4 and a $T$-depth of 2 \cite{gidney_halving_2018}. The overall $T$-count of the controlled SEL circuit is then $4L-4$ where $L$ is the number of Pauli strings in $H$. Moreover, the $T$-depth is $2L-2$ and the procedure introduces $\lceil \log_2 L \rceil$ additional ancillas.

To implement the controlled PREP circuit, a \ac{QROM} is used to load classical data (i.e., coefficients $c_i$) into a quantum computer \cite{morisaki2024classical, babbush_encoding_2018}. This can significantly reduce the $T$-complexity compared to naive implementations. The goal is to implement a transformation 
\begin{align}
    \ket{0}^{1+2\mu+2\lceil \log_2 L \rceil} \to \sum_{i=1}^{L} \sqrt{\frac{\tilde{c_i}}{\lambda}}  \ket{i} \ket{\mathrm{temp}_i}.
    \label{eq_app:prep_state}
\end{align}
Here, $\tilde{c_i}/ \lambda$ are $\mu$-bit approximations of the exact probabilities $|c_i|/\lambda$, and $\ket{\mathrm{temp}_i}$ a temporary junk register that is approximately uncomputed with $\prep^{\dagger}$. The coefficients only need to be implemented accurately enough to ensure that the final energy error remains below a target threshold $\Delta E$. As we shall see in \Eq \ \eqref{eq_app:energy_error_qpe_prep}, a preparation error of $\epsilon_{\mathrm{PREP}} \leq \Delta E / 2 \lambda$ is sufficient to achieve the desired energy error. With that, the error in the prepared coefficients can be estimated as
\begin{align}
    \frac{|c_i - \tilde{c}_i|}{\lambda} \leq \frac{1}{2^\mu L}
    \approx \frac{\Delta E}{2L \lambda\left(1+ \frac{\Delta E^2}{4 \lambda^2} \right)},
\end{align}
which is derived in the Appendix of Ref. \cite{babbush_encoding_2018}. Therefore, we choose
\begin{align}
    \mu =  \left \lceil \log_2\left(\frac{2 \lambda}{\Delta E} \right) + \log_2 \left(1+\frac{\Delta E^2}{4 \lambda^2} \right) \right \rceil.
\end{align}

To prepare the state in \Eq \ \eqref{eq_app:prep_state}, first an equal superposition state is prepared over $L$ computational basis states. Using the \ac{QROM}, precomputed probability values (keep$_i$ and alternate$_i$) are loaded to selectively retain or swap an index. This process redistributes the amplitudes, ultimately producing a quantum state with the correct probabilities that correspond to the coefficients. The procedure requires $1 +2\mu + \lceil \log_2 L \rceil$ ancillas next to $\lceil \log_2 L \rceil$ work qubits and has an overall $T$-count of $4(L + \mu) +  2k + 10 \lceil\log_2 J \rceil$ with $L = 2^kJ$. Moreover, the $T$-depth is $2(L + \mu) +  2k + 6 \lceil\log_2 J \rceil$. Here, contributions smaller than the specified terms are omitted. For details, we refer to Ref. \cite{babbush_encoding_2018}, Section III.D. Note that for Hamiltonians with diagonal Coulomb operators, the $T$-complexity can further be reduced, however, we only consider general Pauli Hamiltonians within this work.

Next to PSP$_H$, the qubitization operator consists of the reflection $2|\mathbf{0}\rangle \langle \mathbf{0}| - \mathbb{I}$. This reflection can be implemented using a C$^m$NOT gate, where $m=\lceil \log_2 L \rceil$, together with Clifford operations. The controlled multi-qubit CNOT gate can be decomposed into a universal gate set using $4 (m +1) - 6$ $T$-gates and $m-1$ ancillas with a $T$-depth of $\sim5$ for $m\leq50$ \cite{nakanishi_systematic_2024}. 

Finally, we need to determine how often the qubitization unitary $Q$ has to be applied to reach a target error in the GSE, $\Delta E$. In addition to the discretization error due to the finite number of bits in the phase estimate, errors introduced by the PREP unitary contribute to the final phase error \cite{babbush_encoding_2018}. Since $\theta_{\mathrm{GS}} = 2 \pi \phi_{\mathrm{GS}}$, we approximate the total error in the phase as
\begin{align}
    \Delta \theta = \sqrt{\left(\frac{2\pi}{2^m} \right)^2 + \epsilon_{\mathrm{PREP}}^2}.
\end{align}
Given that $\exGSE = \lambda \cos(2\pi \phi_{\mathrm{GS}}) = \lambda \cos(\theta_{\mathrm{GS}})$, the energy error is roughly
\begin{align}
    \Delta E = \lambda \Delta \cos(\theta_{\mathrm{GS}}) \leq \lambda \Delta \theta \approx \lambda \sqrt{\left(\frac{2\pi}{2^m} \right)^2 + \epsilon_{\mathrm{PREP}}^2}.
    \label{eq_app:energy_error_qpe_prep}
\end{align}
Hence, we can choose
\begin{align}
    m = \left\lceil \log_2 \left( \frac{4 \pi \lambda}{\Delta E} \right) \right\rceil < \log_2 \left( \frac{8 \pi \lambda}{\Delta E} \right)
\end{align}
and 
\begin{align}
    \epsilon_{\mathrm{PREP}} \leq \frac{\Delta E}{2 \lambda},
\end{align}
such that both contributions to the phase error are at most $\Delta E/(2\lambda)$ and the total error in \Eq \ \eqref{eq_app:energy_error_qpe_prep} is $< \Delta E$. With that, we need $2^m - 1 \approx 2^m$ applications of $Q$. The total $T$-count $N_T$ is thus estimated as
\begin{align}
    N_{T} \approx 2^m (2 \cdot [4(L +\mu) +  2k + 10 \lceil\log_2 J \rceil]  + 4L - 4 \notag \\ + 4 (\lceil \log_2 L \rceil +1) - 6).
\end{align}
Moreover, the $T$-depth $d_T$ is given by
\begin{align}
    d_{T} \approx 2^m (2 \cdot [2(L +\mu) +  2k + 6 \lceil\log_2 J \rceil]  + 2L - 2 + 5).
\end{align}
Additionally, we add the $T$-count and the $T$-depth of the respective ansatz and the shots estimated with a majority vote for each iteration to $N_{T}$ and $d_T$, respectively.

\bibliographystyle{apsrev4-2}
\bibliography{literature.bib}

\end{document}